\documentclass[preprint2]{aastex}
\input psfig.sty

\newcommand{\ee}{$e^\pm$}

\newcommand{\sax}{{\it Beppo\-SAX}}

\newcommand{\xte}{{\it RXTE}}
\newcommand{\pca}{{\it RXTE}/PCA}
\newcommand{\ginga}{{\it Ginga}}
\newcommand{\asca}{{\it ASCA}}
\newcommand{\gro}{{\it CGRO}}

\begin{document}

\title{Understanding the long-term spectral variability of
Cygnus X-1 with BATSE and ASM observations}

\author{Andrzej A. Zdziarski\altaffilmark{1}, Juri Poutanen\altaffilmark{2,3},
William S. Paciesas\altaffilmark{4} and Linqing Wen\altaffilmark{5}
}

\altaffiltext{1}{Centrum Astronomiczne im.\ M. Kopernika, Bartycka 18,
00-716 Warszawa, Poland; aaz@camk.edu.pl}

\altaffiltext{2}{Astronomy Division, P.O.Box 3000, 90014 University of Oulu,
Finland; juri.poutanen@oulu.fi}

\altaffiltext{3}{Stockholm Observatory, 10691 Stockholm, Sweden}

\altaffiltext{4}{Department of Physics, National Space Science and Technology
Center, University of Alabama in Huntsville, Huntsville, AL 35899, USA;
Bill.Paciesas@msfc.nasa.gov}

\altaffiltext{5}{Center for Space Research, Massachusetts Institute of
Technology, Cambridge, MA 02139; lwen@ligo.caltech.edu}

\begin{abstract} We present a comprehensive analysis of all observations of 
Cygnus X-1 by the \gro/BATSE (20--300 keV) and by \xte/ASM (1.5--12 keV) until 
2002 June, including $\sim 1200$ days of simultaneous data. We find a number of 
correlations between fluxes and hardnesses in different energy bands. In the 
hard (low) spectral state, there is a negative correlation between the ASM 
1.5--12 keV flux and the hardness at any energy. In the soft (high) spectral 
state, the ASM flux is positively correlated with the ASM hardness but 
uncorrelated with the BATSE hardness. In both spectral states, the BATSE 
hardness correlates with the flux above 100 keV, while it shows no correlation 
with the 20--100 keV flux. At the same time, there is clear correlation between 
the BATSE fluxes below and above 100 keV. In the hard state, most of the 
variability can be explained by softening the overall spectrum with a pivot at 
$\sim 50$ keV. There is also another, independent variability pattern of lower 
amplitude where the spectral shape does not change when  the luminosity changes. 
In the soft state, the variability is mostly caused by a variable hard 
(Comptonized) spectral component  of a constant shape superimposed on a constant 
soft blackbody component. These variability patterns are in agreement with the 
dependencies of the rms variability on the photon energy in the two states.

We also study in detail recent soft states from late 2000 till 2002. The last of 
them has lasted so far for $> 200$ days. Their spectra are generally harder in 
the 1.5--5 keV band and similar or softer in the 3--12 keV band than the spectra 
of the 1996 soft state whereas the rms variability is stronger in all the ASM 
bands. On the other hand, the 1994 soft state transition observed by BATSE 
appears very similar to the 1996 one.

We interpret the variability patterns in terms of theoretical Comptonization 
models. In the hard state, the variability appears to be driven mostly by 
changing flux in seed photons Comptonized in a hot thermal plasma cloud with an 
approximately constant power supply. In the soft state, the variability is 
consistent with flares of hybrid, thermal/nonthermal, plasma with variable power 
above a stable cold disk. The spectral and timing differences between the 1996 
and 2000--02 soft states are explained by a decrease of the color disk 
temperature. Also, based on broad-band pointed observations simultaneous with 
those of the ASM and BATSE, we find the intrinsic bolometric luminosity 
increases by a factor of $\sim 3$--4 from the hard state to the soft one, which 
supports models of the state transition based on a change of the accretion rate. 
\end{abstract}


\section{Introduction}
\label{intro}

Cygnus X-1 is an archetypical black-hole X-ray binary discovered in 1964 (Bowyer
et al.\ 1965). Due to its brightness and the persistency of emission, its
observational and theoretical studies have been of great importance for
understanding of the process of accretion onto black holes in general.

A specific important tool for testing theoretical models is the study of
spectral variability. Pointed observations of Cygnus X-1 with X-ray and
$\gamma$-ray telescopes have provided detailed information about its broad-band
spectra (Gierli\'nski et al.\ 1997, hereafter G97, Gierli\'nski et al.\ 1999,
hereafter G99, Di Salvo et al.\ 2001, hereafter D01, Frontera et al.\ 2001,
hereafter F01, McConnell et al.\ 2002, hereafter M02), and its short time scale
variability (e.g., Revnivtsev, Gilfanov, \& Churazov 2000; Maccarone, Coppi, \&
Poutanen 2000). However, many black hole sources, including Cygnus X-1, also
show dramatic spectral changes such as spectral transitions which occur on the
time scales of weeks. Understanding the causes of these variations may require
long uninterupted observations which are not available. The only observationally
possible alternative is to monitor a source by all-sky monitors such as the ASM
detectors aboard \xte\/ and the BATSE detectors aboard \gro.

Cygnus X-1 is an excellent target for such studies. It has been monitored by the 
ASM since 1996, and by the BATSE from 1991 to 2000. During almost 5 years 
(1996--2000) the monitoring was simultaneously by both detectors. Spectral 
variability in the ASM data alone have been studied by Wen, Cui, \& Bradt 
(2001), Reig, Papadakis, \& Kylafis (2002) and Smith, Heindl, \& Swank (2002), 
and some aspects of spectral variability in the BATSE data have been studied by 
Crary et al.\ (1996). Some partial analyses of joint ASM-BATSE variability have 
been done by Zhang et al.\ (1997, hereafter Z97) for the 1996 state transition, 
and by Brocksopp et al.\ (1999a). Here we present a comprehensive joint analysis 
of spectral variability in the full currently available ASM data and of all of 
the BATSE data.

\section{Data analysis}
\label{data}

To ensure both a sufficient overlap between the observations by ASM and by BATSE 
and sufficient statistics, we have chosen to use data binned on one-day 
timescale. The ASM data (average daily count rates and their standard errors) 
have been taken from the public database at xte.mit.edu/ASM\_lc.html. The data 
analyzed here cover the time interval from 1996 January 5 to 2002 June 6 (MJD 
50087--52431), yielding 2127 days with usable measurements. In order to convert 
counts to energy units, we have constructed our own matrix based on comparison 
with pointed observations taken on the days with the ASM coverage, see Appendix 
A. In this way, we obtain energy fluxes in the photon energy, $E$, ranges of 
1.5--3 keV, 3--5 keV, and 5--12 keV.

For the BATSE data, we generally followed the standard Earth occultation
analysis methodology (Harmon et al.\ 2002). Counting rate spectra from
individual occultation steps were averaged over one-day intervals. Energy fluxes
in the 20--100 keV and 100--300 keV ranges were derived for each day by fitting
a power-law spectrum (with variable photon spectral index, $\Gamma$, defined by
the photon flux  $\propto E^{-\Gamma}$) separately to the hardware channel
ranges that corresponded most closely to the chosen energy intervals. To
minimize interference from Cygnus X-3, we excluded data from days on which the
angle between Cygnus X-3 and the closest point on the Earth's limb at the time
of Cygnus X-1 occultation was $<2.5\degr$. We also excluded days on which fewer
than four occultation steps were measured, since these are more likely to be
affected by systematic errors.

The above procedure yields 2729 days with usable data from 1991 April 25 to
2000 May 22 (MJD 48371--51686). In Appendix A, we compare the BATSE data with
those from the same pointed measurements as used for the ASM, and find them to
be in a good general agreement.

The data sets above include 1187 days of usable data with simultaneous ASM-BATSE
coverage. Using these data, we study here correlations between the ASM and BATSE
fluxes and spectral slopes.

Often, hardness of spectra is described by the ratio of count rates or fluxes,
$R$, in two adjacent channels. However, even if energy fluxes are used, this
value depends on the width of the channels. To describe the hardness in a more
objective way, we define here the effective photon power-law spectral index,
$\Gamma$, by the condition of the ratio of the energy fluxes in the power law
spectrum, $R$, being equal to that in the observed spectrum. This index is
analogous to colors used in stellar astrophysics. When two adjacent channels
have boundary energies of $E_1< E_2< E_3$, the relationship between $\Gamma$ and
$R$ is given by,
\begin{equation} R = \cases{ {E_3^{2-\Gamma} -
E_2^{2-\Gamma}\over E_2^{2-\Gamma} - E_1^{2-\Gamma}}, & $\Gamma\neq 2$;\cr
{\ln{E_3/E_2}\over \ln{E_2/E_1}}, & $\Gamma= 2$.\cr} \label{eq:hr}
\end{equation}
The uncertainty of $\Gamma$ is given by the propagation of
errors, $\Delta \Gamma = \Delta R / \vert {\rm d} R/{\rm d} \Gamma\vert$ where
$\Delta R$ is the uncertainty of $R$. This allows us to obtain two effective
indices for the ASM data, and one for BATSE. The BATSE effective index has a
substantially better accuracy than either of the fitted BATSE indices.

\section{Lightcurves}
\label{light}

Figure \ref{lc1}a presents the 1.5--12 keV ASM and 20--300 keV BATSE energy flux
light curves. We have rebinned the original ASM and BATSE one-day average data
to assure that the error on each flux point is $<10\%$ and $<20\%$,
respectively. This reduces the number of data points by only 5 and 7,
respectively, and the relative accuracy of most of the values in Figure
\ref{lc1}a is much better than the above ones. Figure \ref{lc1}b shows the
corresponding hardness in the 3--12 keV and 20--300 keV bands, as given by the
effective index of equation (\ref{eq:hr}). We have chosen here the 3--12 keV
index in order to avoid the effect of variable absorption by the line-of-sight
matter, which affects mostly the spectrum below 3 keV.

\begin{figure*}
\centerline{\psfig{file=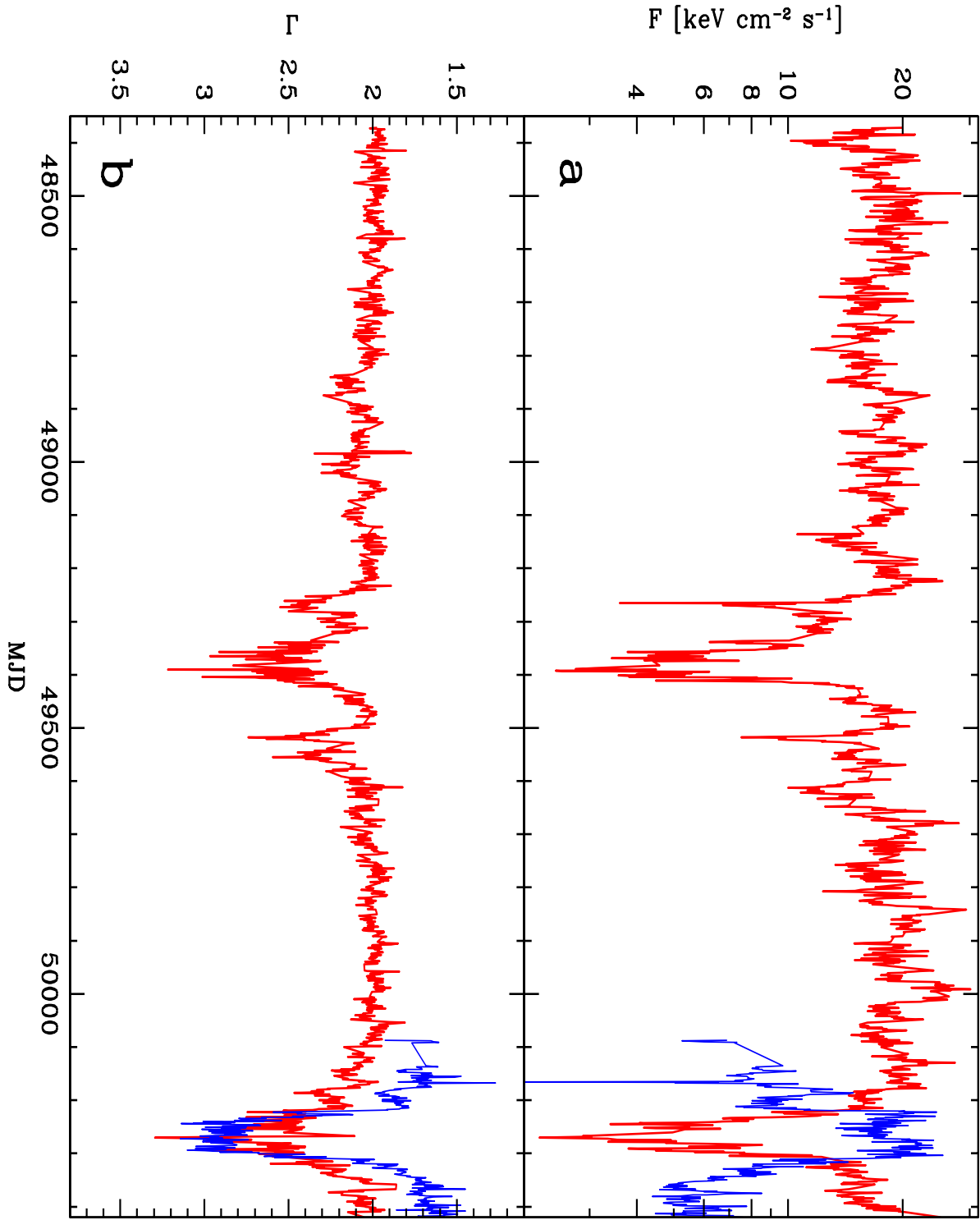,width=16cm,height=20cm,angle=-90
}}
\caption{(a) Light curves of Cygnus X-1 as observed by the \xte/ASM and
\gro/BATSE. Thin blue and heavy red lines give the energy flux in the 1.5--12
keV (ASM) and 20--300 keV (BATSE) bands, respectively. (b) The corresponding
time dependencies of the effective 3--12 keV (blue) and 20--300 keV (red) photon
indices. Note occasional gaps in the coverage, joined by straight lines.
\label{lc1} }
\end{figure*}
\setcounter{figure}{0}
\begin{figure*}
\centerline{\psfig{file=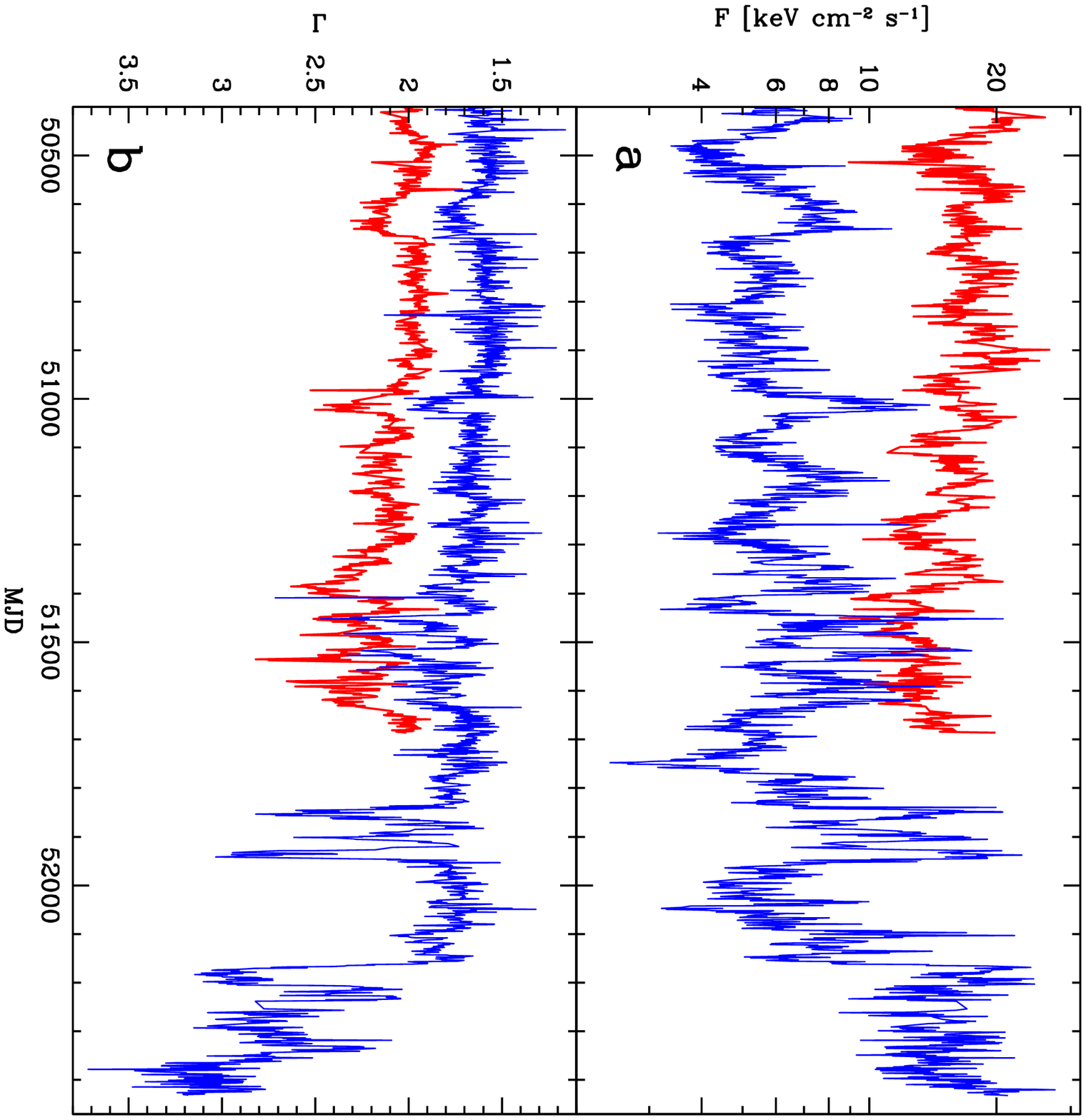,width=16cm,height=20cm,angle=-90
}}
\caption{continued. }
\end{figure*}

\begin{figure*}[t!]
\epsscale{2.0}
\plotone{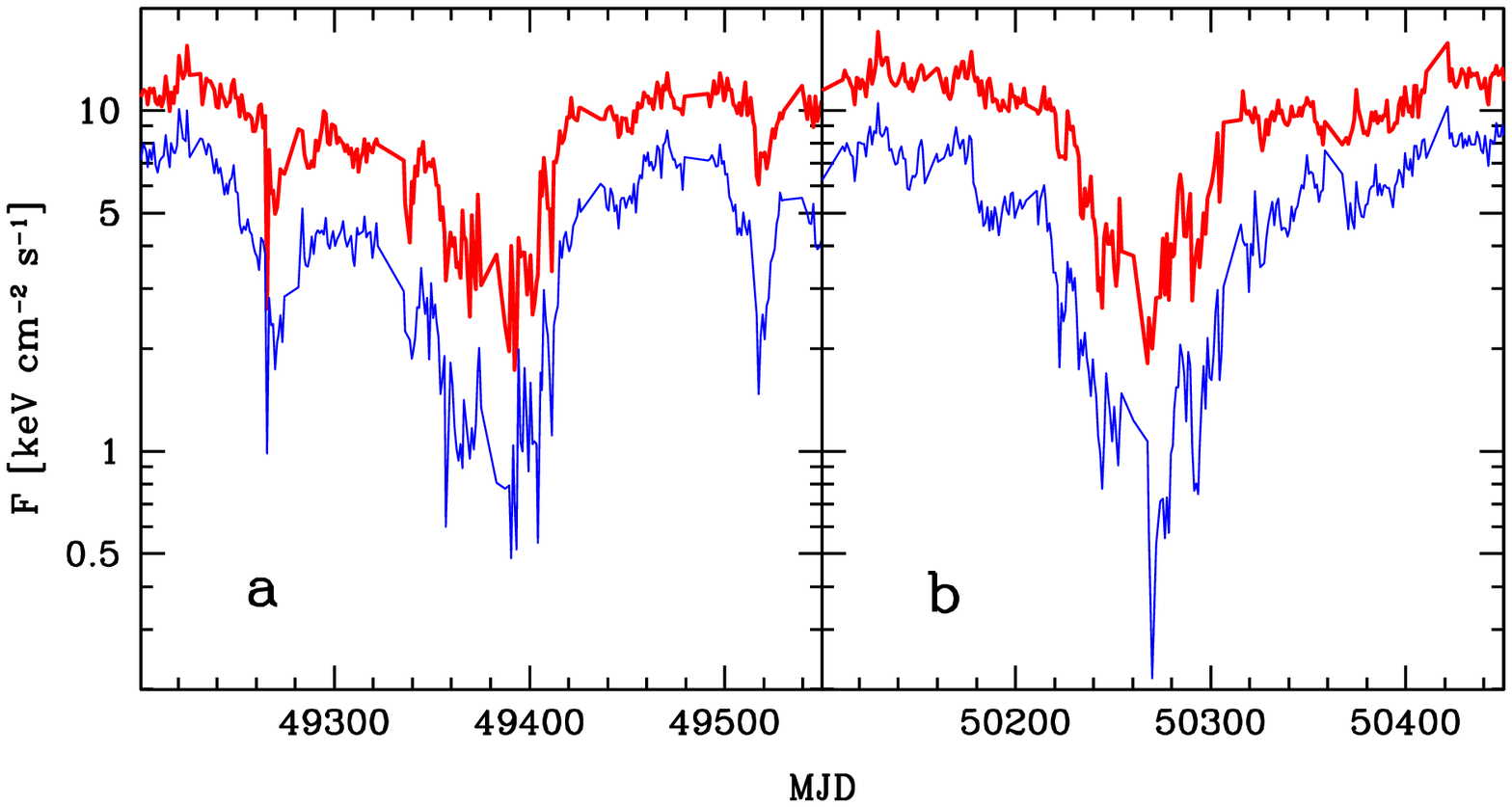}
\caption{Comparison of the state transitions observed by BATSE in (a) 1994 and
(b) 1996. The thick (upper) and thin (lower) curves give the 20-100 keV and
100--300 keV energy fluxes. The two transitions show a striking similarity.
\label{lc_batse} }
\end{figure*}

In Figure \ref{lc1}, we see a soft-state transition in 1994, observed by both
BATSE and OSSE (e.g., G99), very similar in form to the famous state transition
of the year 1996 (Z97), as illustrated in more detail in Figure \ref{lc_batse}.
Unfortunately, Cygnus X-1 was not observed at lower energies during this
transition.

During the simultaneous operation of ASM and BATSE, we see the well-known 
overall anticorrelation of the fluxes from the two instruments.  Interestingly, 
the 3--12 keV ASM index was rather similar to the BATSE one during the 1996 soft 
state.  We also in Figure \ref{lc1}a a number of flares in the ASM flux which 
were {\it not\/} accompanied by any significant drops in the BATSE flux.

\begin{figure*}
\centerline{\psfig{file=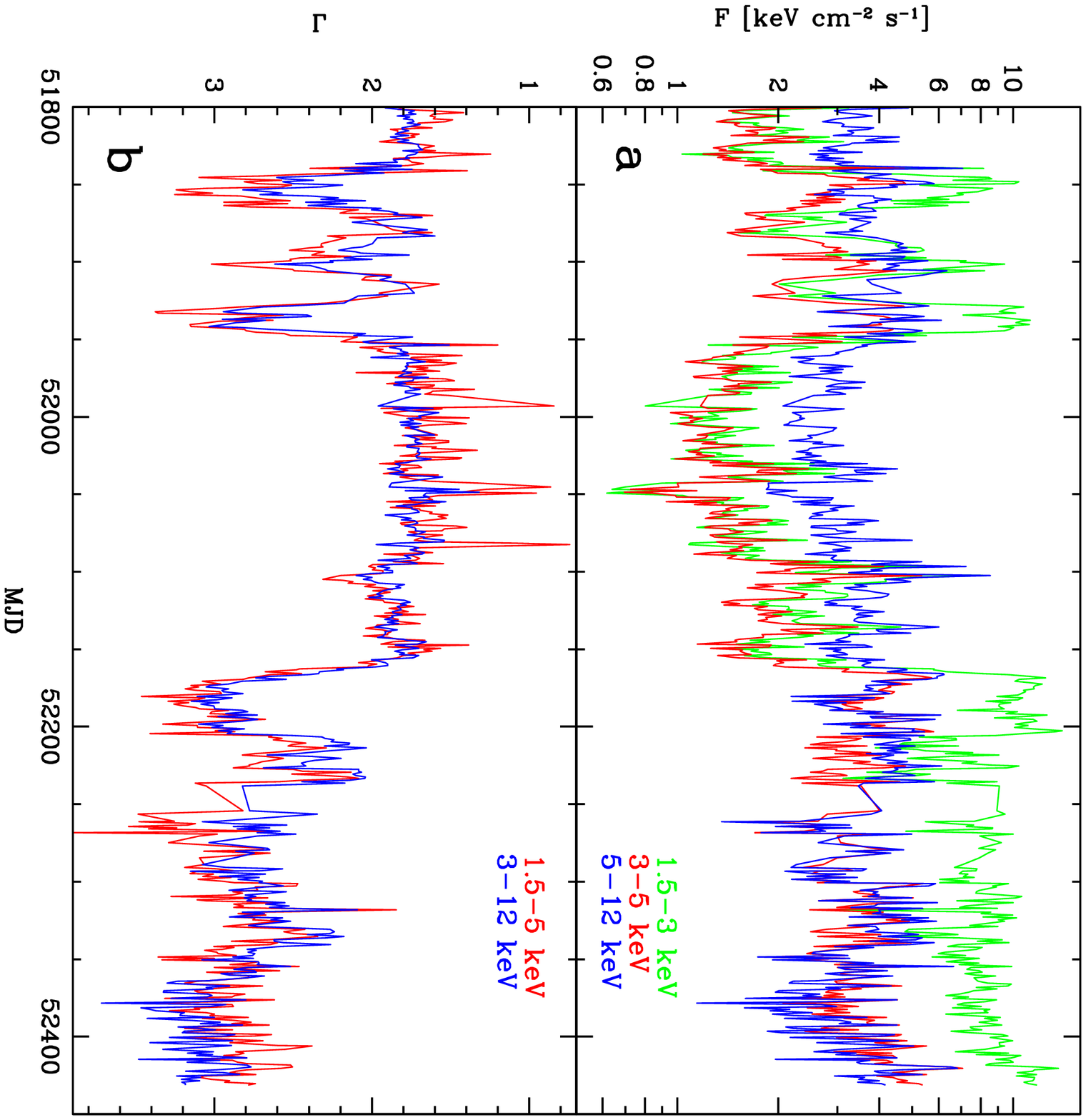,width=16cm,height=20cm,angle=-90
}}
\caption{(a) Recent ASM light curves of Cygnus X-1, with the 1.5--3 keV, 3--5
keV, and 5--12 keV fluxes shown in green, red, and blue lines, respectively. (b)
The corresponding time dependencies of the effective 1.5--3 keV (red) and 3--12
keV (blue) photon indices.
\label{lc_asm} }
\end{figure*}

In addition to the 1996 soft state, we also see three shorter episodes when the 
ASM flux became high and its spectrum very soft, with $\Gamma(3\!-\!\!12\, {\rm 
keV})> 2.4$, around MJD 51850, 51900, and 51940, reported by Cui, Feng, \& 
Ertmer (2002). More recently, another soft state lasting for about two months 
occured around MJD $\sim 52150$--52210. After this, Cygnus X-1 went to an 
intermediate [i.e., still soft, with $\Gamma(3\!-\!\!12\, {\rm keV}) \ga 2$] 
state for about a month. This was followed by yet another occurence of the soft 
state, which started around MJD 52230 and has lasted almost continuously (with 
some short intervals of an intermediate state around MJD 52350) for $\sim 200$ 
days so far, i.e., much longer than other state transitions of Cygnus X-1 
observed so far. These recent state transitions are shown in detail in Figure 
\ref{lc_asm}. It is worth noticing that the spectrum during the last soft state 
was substantially softer than during the 1996 soft state, in the range of 
$\Gamma(3\!-\!\!12\, {\rm keV})\sim 3$--3.5. Very interestingly, the spectrum 
since MJD 52160 until now (MJD 52431), i.e., for $>270$ days, has shown 
$\Gamma(3\!-\!\!12\, {\rm keV}) \ga 2$, i.e., much softer than the usual 
$\Gamma(3\!-\!\!12\, {\rm keV}) \sim 1.7$. Unfortunately, no monitoring at 
higher energies was available after MJD 51686, including those highly 
interesting state transitions.

We note that there have been recent reports of short (with duration $\ll 1$
day), very strong, flares from the direction of Cygnus X-1 measured at photon
energies $\ga 20$ keV, at which range the flux exceeded the typical, hard-state,
flux by a factor of $\sim 10$ (Stern, Beloborodov, \& Poutanen 2001; Golenetskii
et al.\ 2002). These flares occured on MJD 49727, 49801, 51289 and 52329. We
have looked into characteristics of the lightcurves around those days, and have
found, unfortunately, no common characteristic that would allow some insight
into the nature of the flares. For example, the last event occured in a soft
state whereas the three previous ones, in hard states.

\section{Spectral Correlations}
\label{correlations}

\subsection{The ASM data}
\label{asm}

Based on the variability from the ASM, we distinguish here three main spectral
states. We define the soft and hard states by the conditions (see Fig.\
\ref{lc1}) of $\Gamma(3\!-\!\!12\, {\rm keV})> 2.4$, $<2.1$, respectively, and
an intermediate state by $\Gamma(3\!-\!\!12\, {\rm keV})=2.1$--2.4. Our
definition of the soft state corresponds, e.g., to MJD between approximately
50230 and 50307 (1996 May 27--Aug.\ 12). These definitions may differ somewhat
from other ones used in the literature.

Wen et al.\ (1999) found strong correlations of the ASM flux and hardness with 
the orbital phase within the hard state, with a peak of the hardness and a dip 
in the flux around the phase zero (i.e. the companion star in front of the X-ray 
source), see also Brocksopp et al.\ (1999a). This effect appears to be due to 
absorption by a partially ionized wind from the companion. On the other hand, no 
phase modulation has been found in the soft state. In order to distinguish 
between this effect and an intrinsic spectral variability, we separately 
consider the hard state data in the orbital phase between 0.85 and 0.15 and 
outside this phase interval. We use the ephemeris of Brocksopp et al.\ (1999b).

\begin{figure*}[t!]
\epsscale{2.0}\plotone{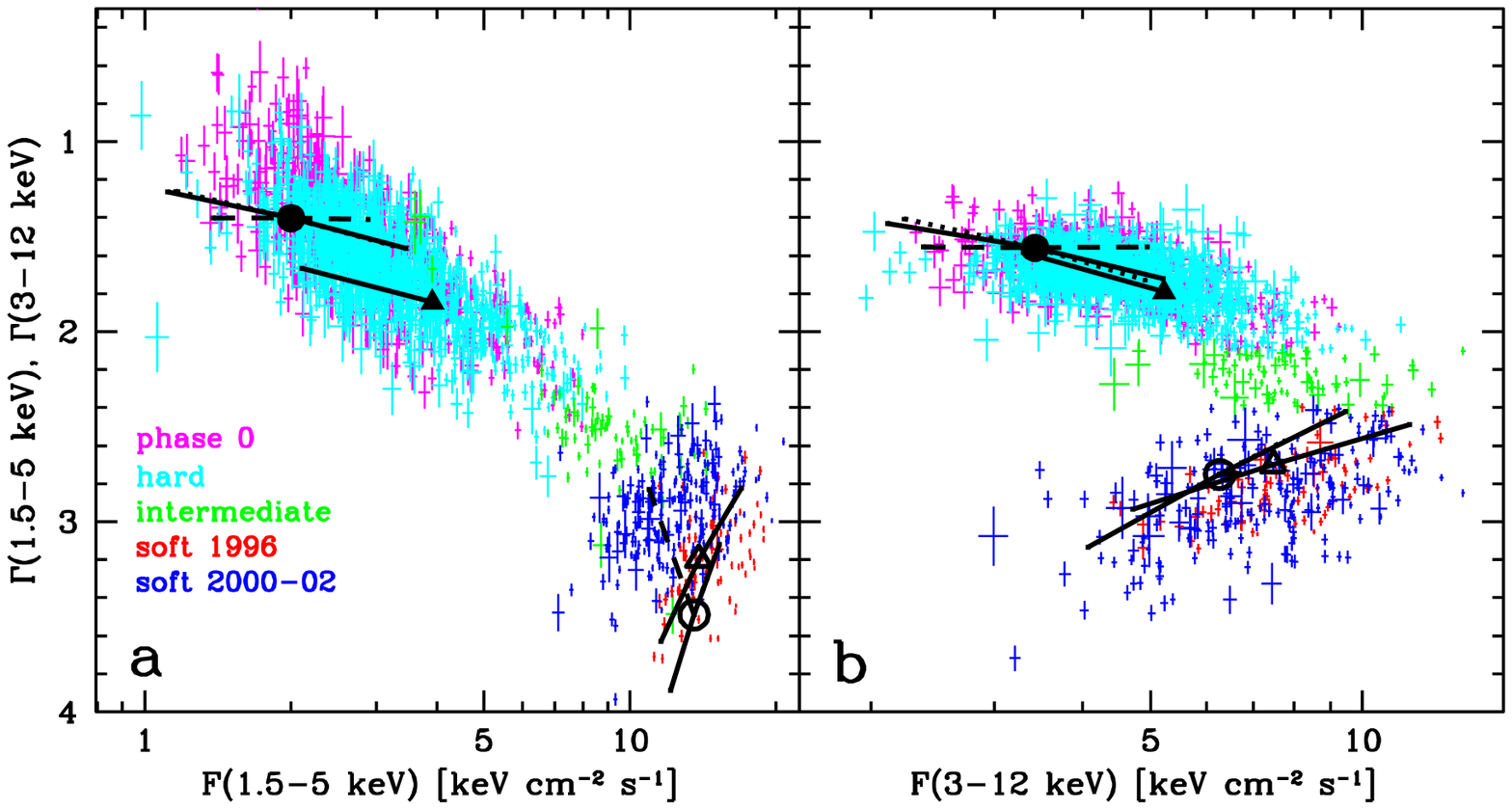}
\caption{The relations between the effective ASM spectral indices [eq.\
(\ref{eq:hr})] and the corresponding fluxes in the energy ranges of (a) 1.5--5
keV and (b) 3--12 keV. Two main spectral states are clearly seen (connected by
an intermediate state), showing the opposite correlations between the hardness
and the flux. The colors red, blue, green, cyan, and magenta correspond to the
1996 soft state, the 2000--02 soft states, the intermediate state
[$\Gamma(3\!-\!\!12\, {\rm keV})= 2.1$--2.4], and the hard state outside and
inside the 0.85--0.15 phase, respectively. The open triangles and circles, and
filled triangles and circles correspond to four spectra from pointed
observations from 1996 May 30--31, June 22, September 12, 1998 May 3--4,
respectively (\S \ref{spectra}). Theoretical variability patterns (\S
\ref{interpretation}) are shown hereafter by black solid lines (hard state:
constant $L_{\rm hard}$, variable $L_{\rm soft}$; soft state: constant $L_{\rm
soft}$, variable $L_{\rm hard}$), dashed lines (hard state: variable total $L$;
soft state: variable temperature of seed photons), and the dotted lines (hard
state: \ee\ pair plasma; soft state: variable $\gamma_{\rm max}$). Some lines on
some figures, e.g., the dotted lines in this figure, happen to almost coincide
with other lines, and are thus poorly visible.
\label{a_flux_index} }
\end{figure*}

We find the logarithms of the fluxes in the three ASM channels strongly 
correlate with each other, linearly within each state. However, the  
coefficients for each of the linear dependence are different in each of the 
three states, as well as the ones for the 1996 soft state are different than 
those for the 2000--02 soft states. This type of variability is most 
conveniently shown on flux-index diagrams (analogous to color-magnitude diagrams 
in stellar astrophysics), Figures \ref{a_flux_index}a, b.

In Figures \ref{a_flux_index}a, b, we see two distinct branches: one formed by
the hard state, and the other one corresponding to the soft states, connected by
the intermediate state. The hard state is slightly extended towards low fluxes
and hard spectra due to local absorption close to the phase zero. Since the
bound-free opacity of cosmic medium generally decreases towards increasing
energies, the phase effect on the index and flux in the 3--12 keV range is
significantly weaker than on those in the 1.5--5 keV one. However, it is clear
that absorption can by no means explain the very strong hardness-flux
anticorrelation in the hard state. A similar hardness-flux anticorrelation is
commonly seen in Seyferts (e.g., Nandra et al.\ 2000), including some radio-loud
ones (e.g., Zdziarski \& Grandi 2001).

We note that the negative correlation in the 1.5--5 keV range is stronger than
that in the 3--12 keV range, but the latter correlation is still relatively
strong. This behavior is consistent with the spectrum pivoting at an energy
above the ASM range. As we find by analyzing the BATSE data, the pivot point is
at an energy between 20 and 100 keV. On the other hand, the horizontal spread in
Figures \ref{a_flux_index}a, b corresponds to a second variability pattern, in
which the X-ray spectrum simply moves up and down without changing its shape.
These two variability patterns are explained theoretically in \S \ref{hard}.

On the other hand, the hardness-flux correlation is positive in the soft states,
as found before by Wen et al.\ (2001). A simple interpretation of this type of
variability is a two-component spectrum, with a constant component at low
energies (e.g., a blackbody or a disk blackbody) and a variable high-energy
tail. Then, the stronger the tail, the higher the flux and the less of the
blackbody cutoff is observed, making the spectrum harder. A theoretical
interpretation of this behavior is given in \S \ref{soft96}.

We note that the existence of two main branches in the index-flux dependence is
independent of the conversion from counts to energy units. If we plot the
analogous figures for the raw ASM counts, the obtained shapes are similar
to those in Figure \ref{a_flux_index}, see Reig et al.\ (2002).

\begin{figure}[t!]
\epsscale{1.0} \plotone{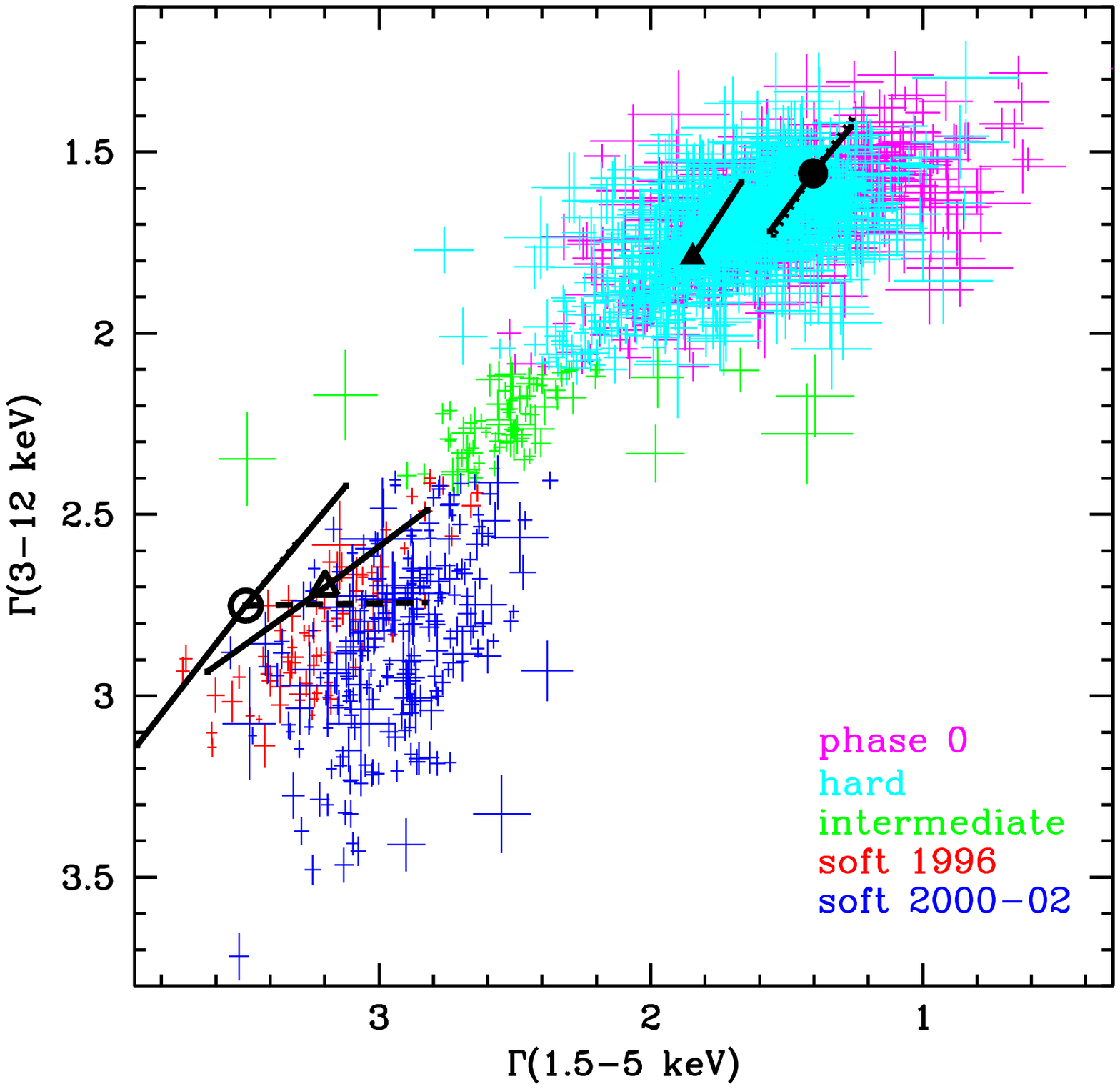}
\caption{The color-color diagram for Cygnus X-1.  The black symbols/lines have
the same meaning as those in Fig.\ \ref{a_flux_index}.
\label{a_c_c} }
\end{figure}

Given the above dependencies, there is a strong linear correlation between the
spectral indices in the 3--12 keV and 1.5--5 keV energy ranges. Figure
\ref{a_c_c} shows a comparison of the two quantities, analogous to color-color
diagrams often shown for X-ray binaries. However, unlike neutron-star binaries
showing changes in the sign of this correlation (e.g., Schulz, Hasinger, \&
Tr\"umper 1989), here the two hardnesses (or $\Gamma$)  are positively
correlated in all states, with only the linear coefficients being
state-dependent. Thus, the X-ray variability of Cygnus X-1 is only on the
diagonal branch of its color-color diagram.

Interestingly, the 2000--02 soft states show distinctly different dependencies 
between the two indices and the 1.5--5 keV index and flux from those for the 
1996 soft state. In Figures  \ref{a_flux_index}a and \ref{a_c_c}, we see that 
$\Gamma(1.5\!-\!\!5\, {\rm keV})$ is generally lower (i.e., the spectrum is 
harder) and the flux is lower at a given hardness during the 2000--02 soft 
states. On the other hand, Figure \ref{a_flux_index}b shows that the 3--12 
keV fluxes and slopes during the two periods of the soft state are rather 
similar, but extending to lower fluxes and softer spectra. Also, the 
flux-hardness correlations in the later soft states remain positive. We discuss 
theoretical explanation of the differences between the two soft states in \S 
\ref{soft02} below.

By inspection of Figure \ref{lc1}, we see that Cygnus X-1 in its hard state can
spend hundreds of days (e.g., MJD 50700--50900) with its X-ray flux changing
only by a factor of two and $\Gamma(3\!-\!\!12\, {\rm keV})$ changing only by
$\pm 0.1$. If a random time interval of 100 days is chosen within the hard
state, the corresponding flux-index points would, in most cases, occupy a
relatively small region within the diagonal hard-state strips in Figures
\ref{a_flux_index}a, b. The relatively large spread of the indices and fluxes
within the hard state seen in these figures is mostly due to the flares and dips
in the long-term X-ray light curve. Thus, although the overall anticorrelation
between the flux and hardness is extremely strong, it is achieved only on very
long time scales, of at least a year. The uniformity of this anticorrelation
also indicates that the minor flares/dips in the light curve still belong to the
hard state, indicating some common underlying physics (see \S \ref{hard}). On
the other hand, flux changes by a factor of two (corresponding to the horizontal
spread of the region covered by the hard-state data in Fig.\ \ref{a_flux_index})
occur on a time scale of tens of days or less. This explains the lack of a
noticeable hardness-flux correlation within time intervals of 5.6 day (i.e., the
orbital period) found by Wen et al.\ (2001).

In the soft states (especially the 1996 one), the correlated flux-index changes
occur on a much shorter time scale of a day or so, as clearly seen in Figure
\ref{lc1}. Thus, although the overall soft-state flux-hardness correlation is
much weaker than the hard-state anticorrelation in Figures \ref{a_flux_index}a,
b, the former can be measured on much shorter time intervals, in agreement with
the very strong correlation within 5.6-day intervals found by Wen et al.\
(2001).

\subsection{The BATSE data}
\label{batse}

\begin{figure*}[t!]
\epsscale{2.0}
\plotone{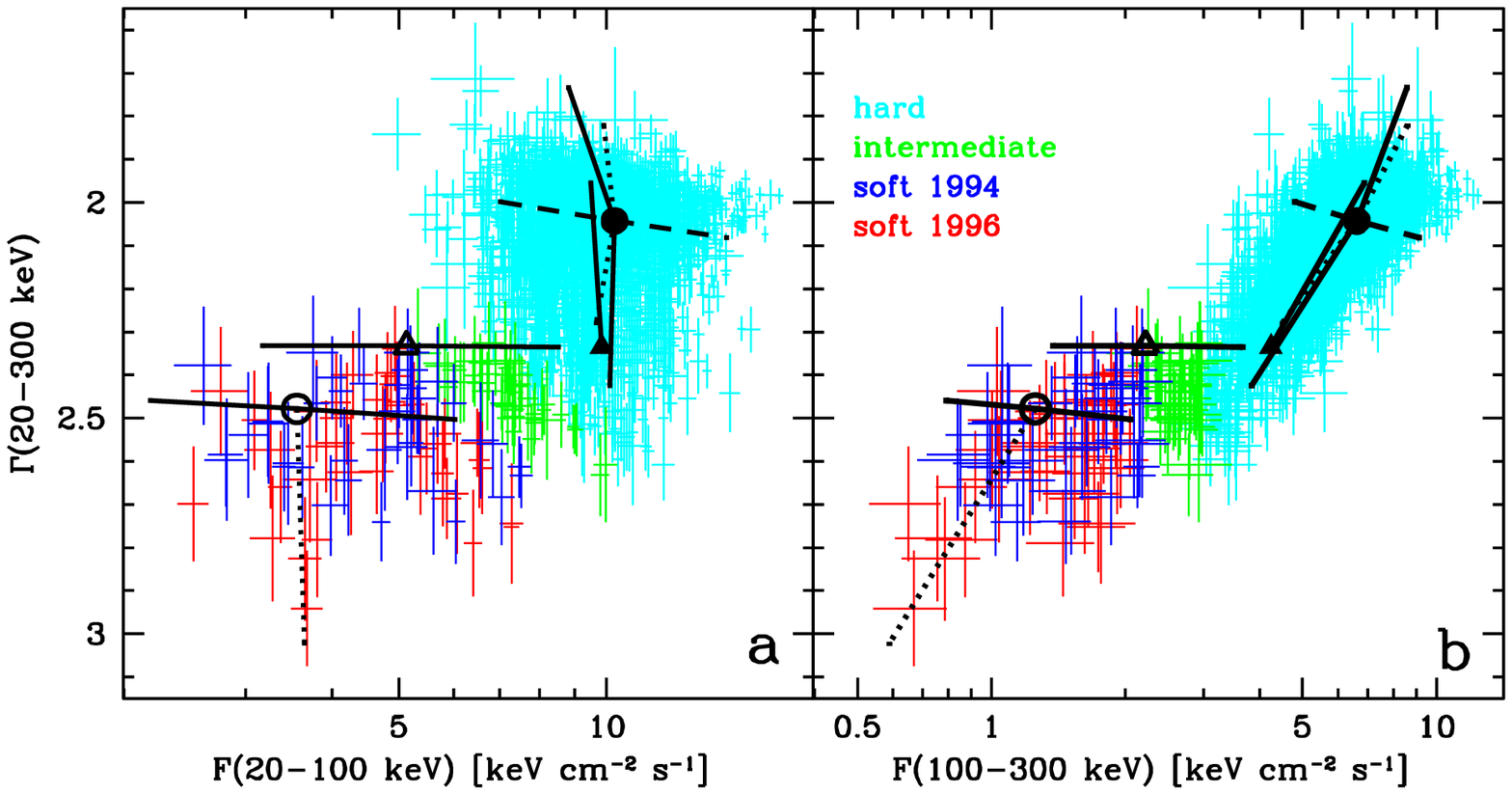}
\caption{The relations between the effective BATSE spectral index [eq.\
(\ref{eq:hr})] and the fluxes in the channels (a) 20--100 keV; (b) 100--300 keV.
The colors blue, red, green, and cyan correspond to the 1994 soft state, the
1996 one, the intermediate state, and the hard state, respectively. The black
symbols and lines have the same meaning as those in Fig.\ \ref{a_flux_index},
except for the dashed lines in the soft state (specific for the 2000--02 state
transitions), which are not shown hereafter.
\label{b_flux_index} }
\end{figure*}

Now, we consider the BATSE data alone. We have found that the spectral state can
be best defined for these data using the 100--300 keV flux (see Fig.\
\ref{lc_batse}). We then define the hard state by $F({\rm
100\!-\!\!300\,keV})>3.0$ keV cm$^{-2}$ s$^{-1}$, and the soft state by $F({\rm
100\!-\!\!300\,keV})<2.2$ keV cm$^{-2}$ s$^{-1}$, with the intermediate state in
between. These definitions are in good agreement with those adopted for the ASM
during the period of the simultaneous ASM-BATSE coverage (e.g., the 1996 soft
state in the BATSE data corresponds to MJD 50232--50305, very similar to the
range obtained from the ASM data in \S \ref{asm}).

Figure \ref{b_flux_index}a shows the effective 20--300 keV spectral index as a
function of the 20-100 keV flux. In the hard state, we still see some
anticorrelation between the hardness and flux, qualitatively similar to those in
the 1.5--12 keV range. However, the anticorrelation is only weak, and, in
particular, much weaker than the one for the 3--12 keV range (which, in turn,
was weaker than that for the 1.5--5 keV range). We then see in Figure
\ref{b_flux_index}b that the 20--300 keV hardness becomes strongly {\it
positively\/} correlated with the 100--300 keV flux. This behavior shows that
the variable spectrum pivots at an energy $\la 100$ keV.

\begin{figure}[t!]
\epsscale{1.0}
\plotone{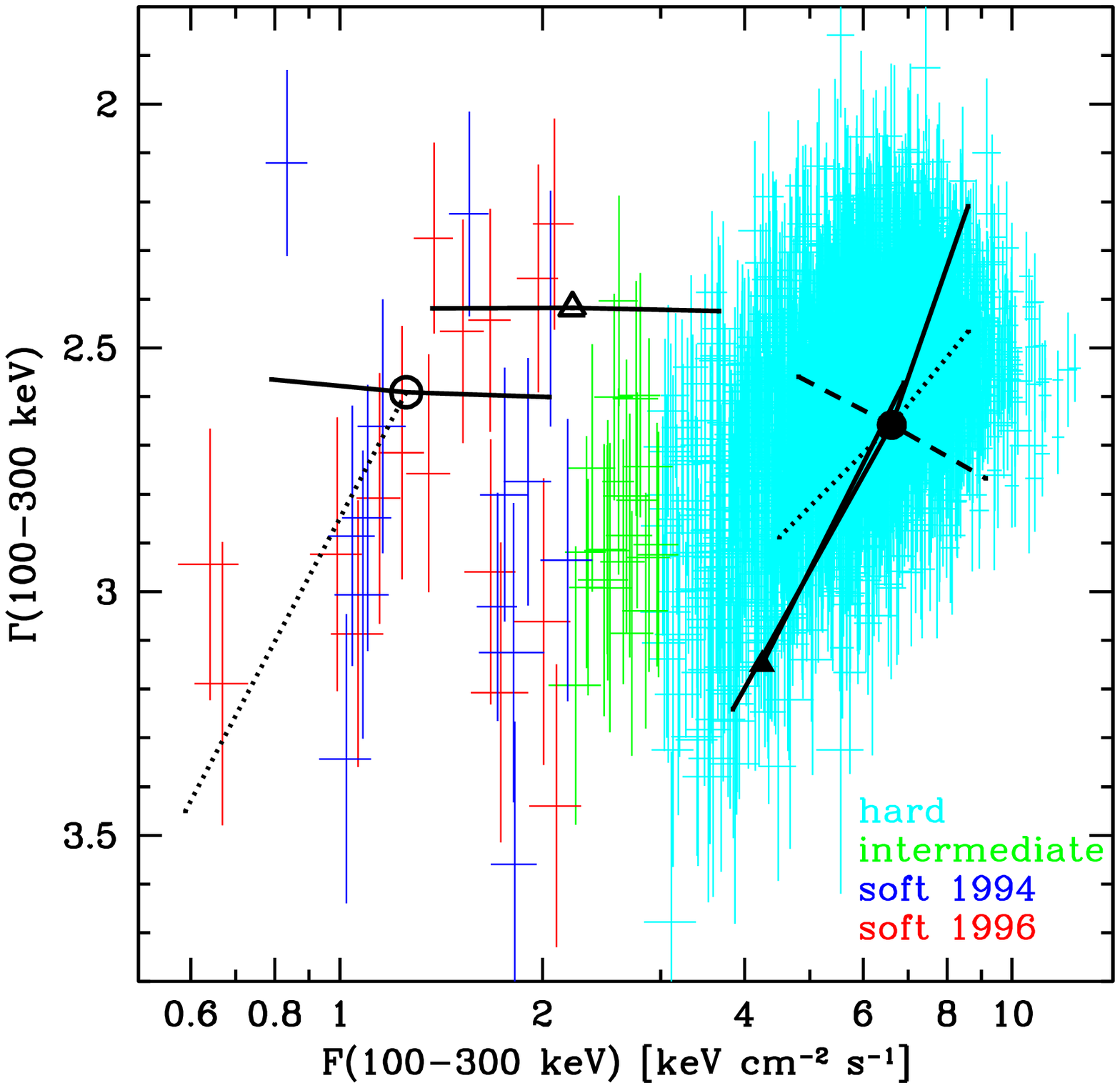}
\caption{The relation between the BATSE spectral index fitted in the 100--300
keV range with the corresponding flux. The black symbols/lines have the same
meaning as those in Fig.\ \ref{a_flux_index}.
\label{b_fitted} }
\end{figure}

The positive hardness-flux correlation in the 100--300 keV range in the hard
state is also seen for the fitted spectral index in that range, as shown in
Figure \ref{b_fitted}. Also, almost no correlation is seen for these quantities
in the 20--100 keV range. This shows that the pivoting effect is not an artifact
of our use of the effective 20--300 keV index. On the other hand, the fitted
indices bear rather large statistical uncertainties. This prevents seeing any
trends in the soft states (Fig.\ \ref{b_fitted}), where the BATSE fluxes are low
and the number of days limited.

Similarly to the case of the ASM data, substantial variations in the range
covered by the source on the $\Gamma$-$F$ diagram during the hard state occur on
long time scales. If, for example, we consider a typical 100-day interval during
the hard state, the corresponding data points would cover a relatively small
part of the cyan region in Figures \ref{b_flux_index}a, b.

\begin{figure}[t!]
\epsscale{1.0}
\plotone{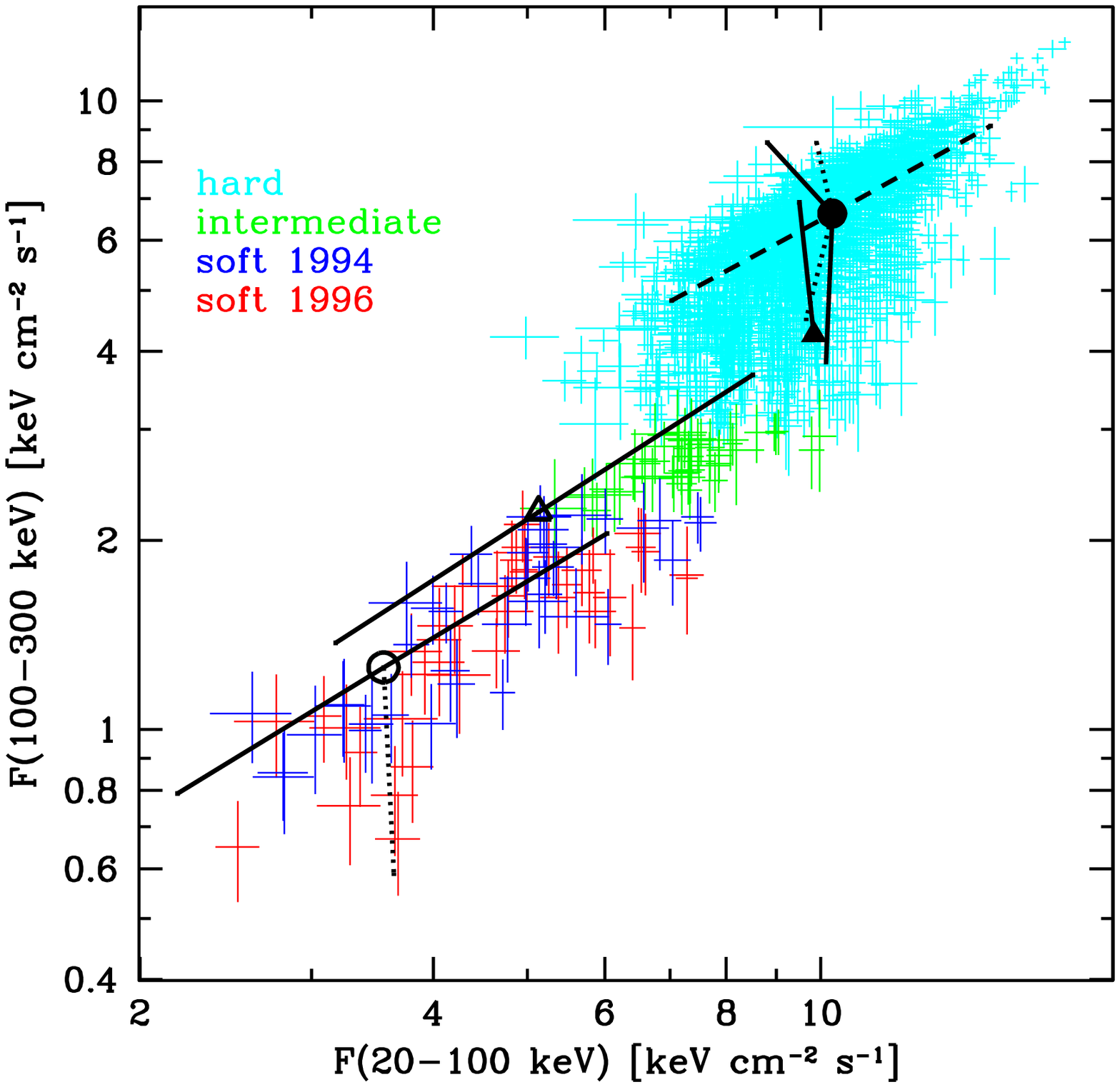}
\caption{The relations between the BATSE fluxes in the 20--100 keV and 100--300
keV energy ranges. The black symbols/lines have the same meaning as those in
Fig.\ \ref{a_flux_index}.
\label{b_flux_flux} }
\end{figure}

Figure \ref{b_flux_flux} shows then the relationship between the 20--100 keV
and 100--300 keV fluxes. We see that, in spite of the pivoting of the spectrum
in the hard state in the BATSE energy range, the two fluxes are still positively
correlated. This appears to be due to two effects. First, the pivot point is
below 100 keV, and a part of the 20--100 keV flux will still positively be
correlated with the 100--300 keV flux. Second, in addition to the pivoting
pattern of variability, there is also variability with the spectrum simply
moving up and down, as discussed in \S \ref{asm}.

We also note that BATSE hardness ratio, $F({\rm 100\!-\!\!300\,keV})/ F({\rm
20\!-\!\!100\,keV})$ [related to the effective spectral index used here via eq.\
(\ref{eq:hr})] shows significant variations by a factor of 3 in the hard state.
This supports earlier findings of Stern et al.\ (2001).

In the soft state, $\Gamma(20\!-\!300\,{\rm keV})$ is only weakly changing and
roughly independent of $F({\rm 20\!-\!\!100\,keV})$, see Figure
\ref{b_flux_index}a. This shows that the 20--100 keV spectrum simply moves up
and down with relatively small changes of the spectral shape, consistent with
our interpretation of the variability in the ASM range being caused by the
amplitude variability of the high-energy tail. In the 20--100 keV range, only
the tail (moving up and down) is observed.  This behavior is also consistent
with the strong positive correlations of the two BATSE fluxes seen in Figure
\ref{b_flux_flux}.

Finally, there appears to be also a weak positive correlation between
$\Gamma(20\!-\!300\,{\rm keV})$ and $F({\rm 100\!-\!\!300\,keV})$ in the 1996
soft state, see Figure \ref{b_flux_index}b. This indicates that the spectrum
above 100 keV softens at low flux levels (as indeed seen directly in the
lightcurve, see Fig.\ \ref{lc_batse}). This points to the existence of a second
variability pattern  in the 1996 soft state, consisting of a softening of the
slope of the tail when the tail amplitude is low.

\subsection{ASM-BATSE correlations}
\label{asm_batse}

\begin{figure}[t!]
\epsscale{1.0}\plotone{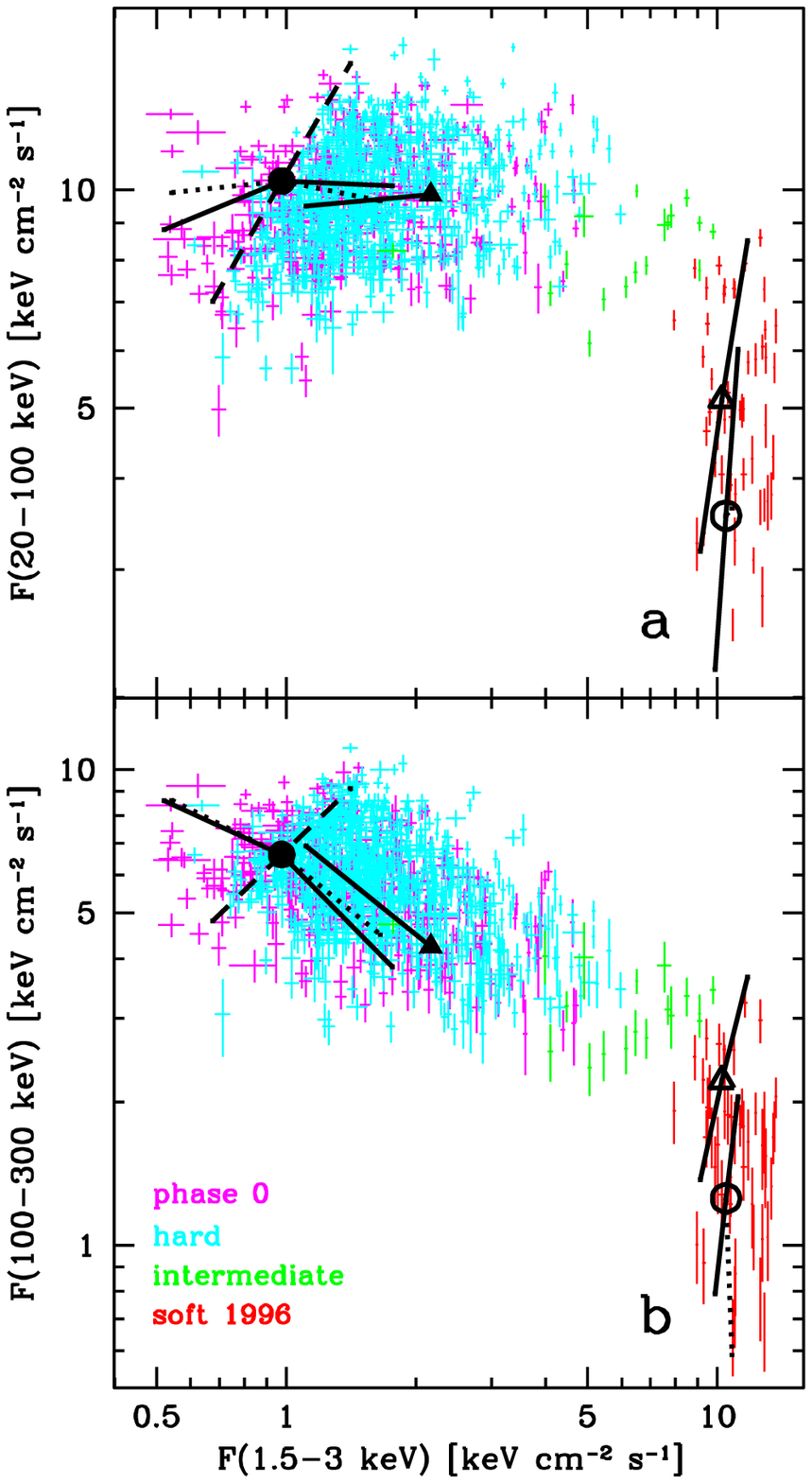}
\caption{The correlation of the ASM 1.5--3 keV flux with the BATSE fluxes:
(a) 20--100 keV, (b) 100--300 keV. The meaning of black symbols/lines
is the same as that in Fig.\ \ref{a_flux_index}.
\label{a_b_flux} }
\end{figure}

We consider now the combined ASM-BATSE data. In general, virtually all
quantities are strongly correlated with each other, in spite of the different
specific times at which the two instruments measure the emission of Cygnus X-1
within a given day. This shows that the intraday variability is relatively weak
compared to the variability on longer time scales. We define here the states
based on $\Gamma(3\!-\!\!12\, {\rm keV})$ (as in \S \ref{asm}).

In Figure \ref{a_b_flux}b, we compare the fluxes in the lowest and the highest
energy channels, i.e., 1.5--3 keV vs.\ 100--300 keV. We see that they are
anticorrelated in the hard state. On the other hand, the 100--300 keV flux is
completely independent of the soft X-ray flux in the soft state. In Figure
\ref{a_b_flux}a, we see that the 20--100 keV flux becomes weakly positively
correlated with the 1.5--3 keV flux in the hard state, while it remains
independent of it in the soft state.

These correlations in the hard state are consistent with the spectrum pivoting
at an energy of $\la 100$ keV. In the soft state, the lack of correlation of the
BATSE fluxes with the 1.5--3 keV one shows that the existence of a constant
component at low energies and a variable and independent component at high
energies. These findings are consistent with those obtained above while
considering the ASM and BATSE data separately.

\begin{figure}[t!]
\epsscale{2.0}\plottwo{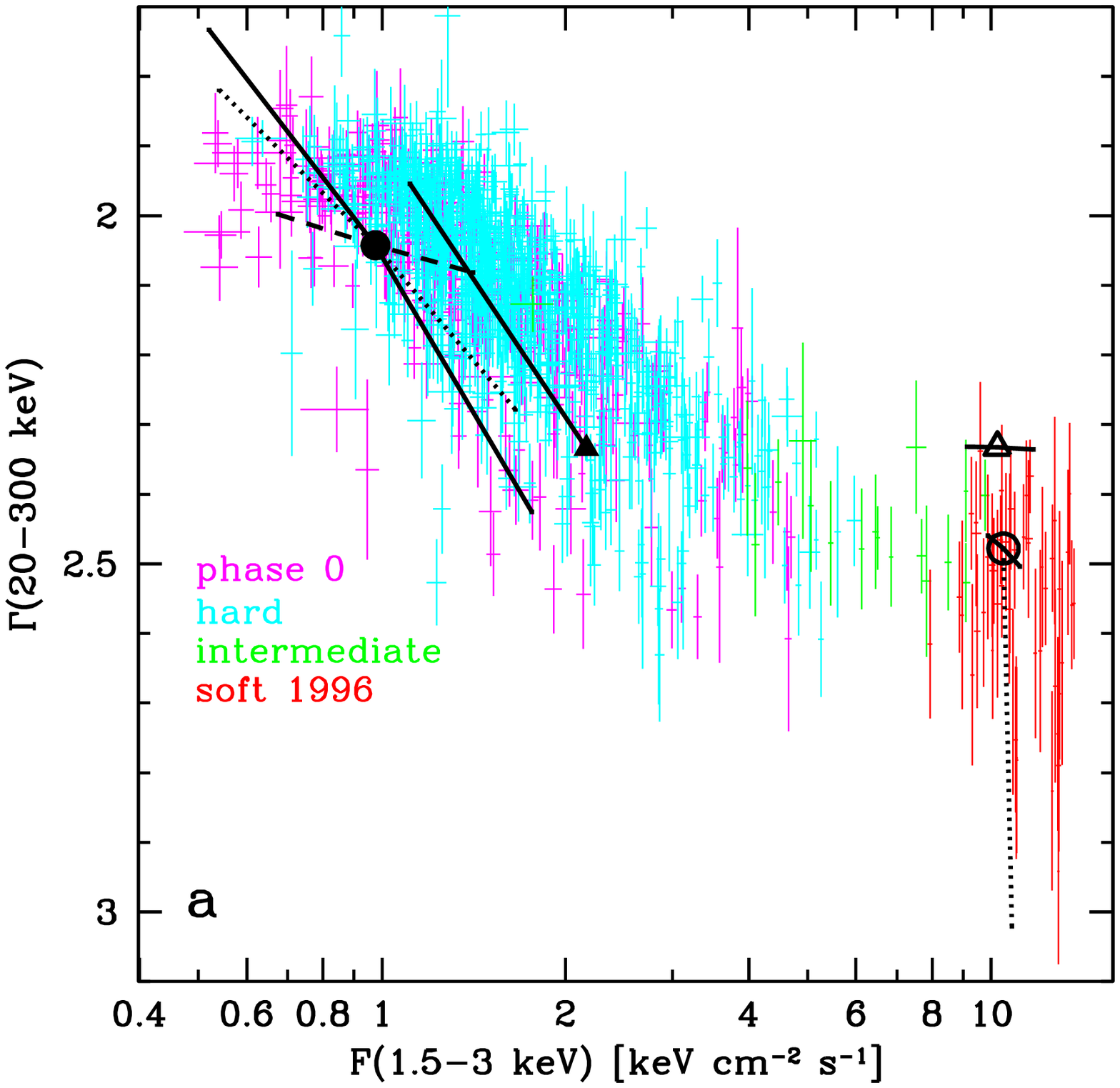}{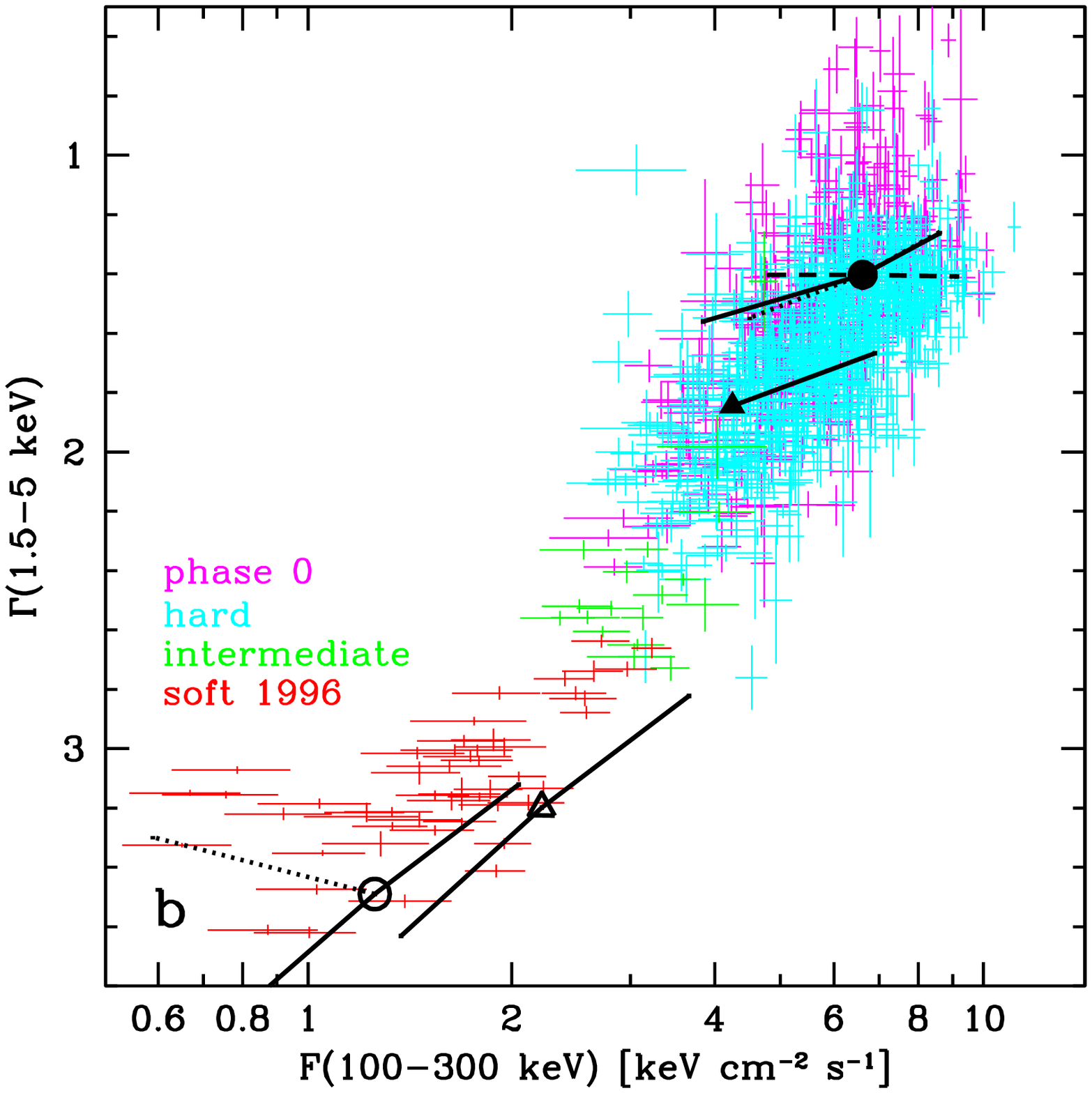}
\caption{ASM-BATSE flux-index correlations. (a) $\Gamma(20\!-\!\!300\,{\rm
keV})$ vs.\  $F(1.5\!-\!\!3\,{\rm keV})$; (b) $\Gamma(1.5\!-\!\!5\,{\rm keV})$
vs.\ $F(100\!-\!\!300\,{\rm keV})$. The meaning of black
symbols/lines is the same as that in Fig.\ \ref{a_flux_index}.
\label{a_b_index} }
\end{figure}

Figures \ref{a_b_index}a, b present the ASM-BATSE flux-index diagrams analogous
to those for Figures \ref{a_flux_index} for the ASM data alone. Figure
\ref{a_b_index}a shows that, in the hard state, the dependence of the BATSE
hardness on the ASM flux is similarly anticorrelated as that of the ASM index on
the ASM flux (Fig.\ \ref{a_flux_index}). In the soft state,
$\Gamma(20\!-\!\!300\,{\rm keV})$ is uncorrelated with the ASM flux, as well as
it shows a substantial spread of values.

On the other hand, the 1.5--5 hardness is now {\it positively\/} correlated in
the hard state with $F(100\!-\!\!300\,{\rm keV})$ (Fig.\ \ref{a_b_index}b),
opposite to the case of the ASM data alone. This is again consistent with
existence of a pivot point below 100 keV, reversing the correlation with respect
to that with a flux at energies $\ll 100$ keV. On the other hand, there is a
moderate hardening of the ASM spectrum with the increasing
$F(100\!-\!\!300\,{\rm keV})$ in the soft state. This is explained by  changing
of the amplitude of the high energy tail with a constant soft, blackbody
component which dominates the ASM flux.

\begin{figure}[t!]
\epsscale{1.0}
\plotone{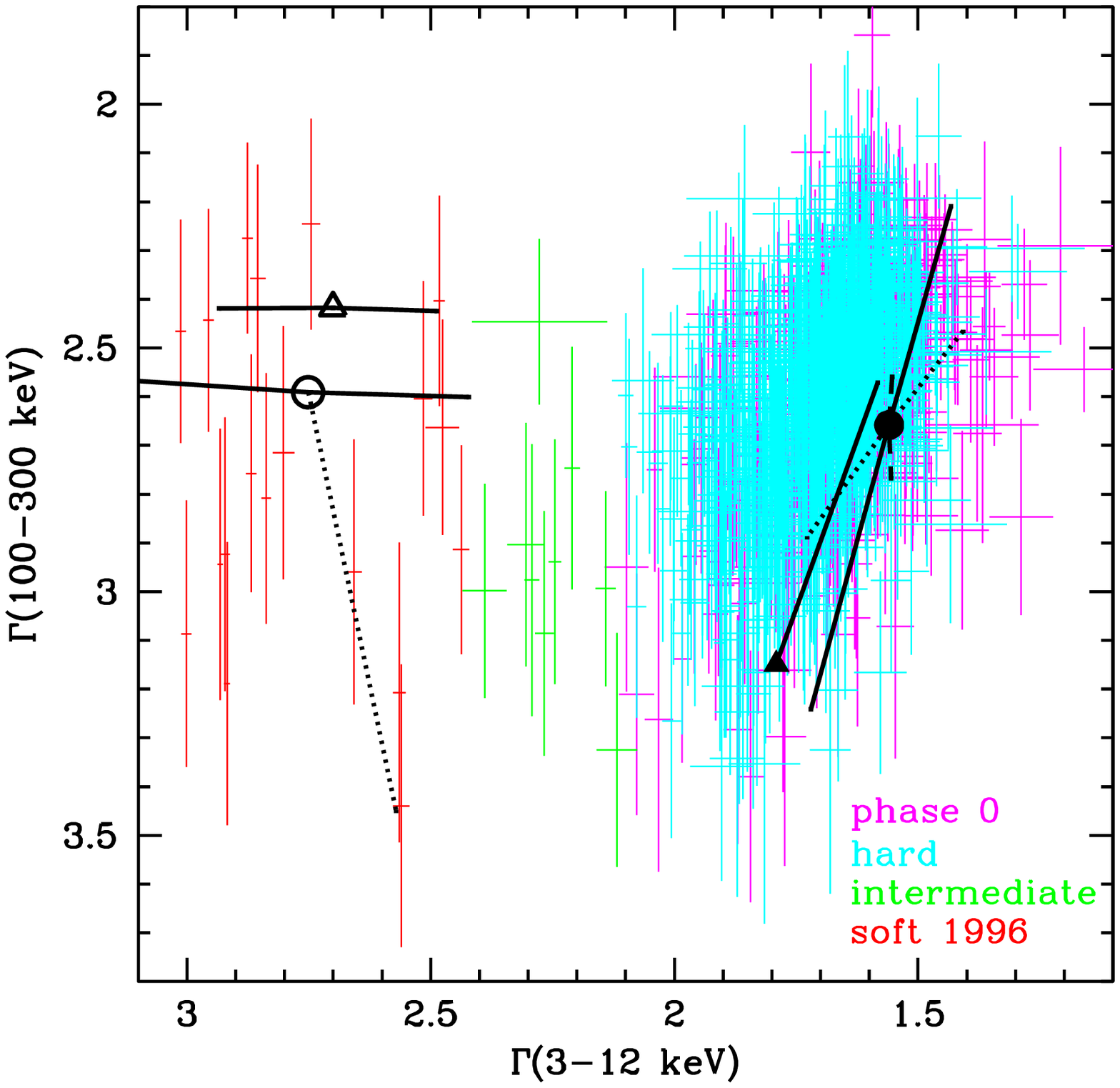}
\caption{The correlation between the 3--12 keV and 100--300 keV indices. The
black symbols/lines have the same meaning as those in Fig.\ \ref{a_flux_index}.
\label{a_b_i_i} }
\end{figure}

Figure \ref{a_b_i_i} shows an ASM-BATSE index-index correlation. The 3--12 keV
hardness is positively correlated with the 100--300 keV one in the hard state,
showing that the spectrum changes its local slope in the same direction at all
measured energies. The BATSE index is substantially softer than that of the ASM,
showing that the spectrum at a given time softens with increasing energy. On the
other hand, there is no correlation in the soft state. This means that
the tail component varies independently of the blackbody one.

\section{Fractional Variability}
\label{rms}

\begin{figure}[t!]
\epsscale{1.0}
\plotone{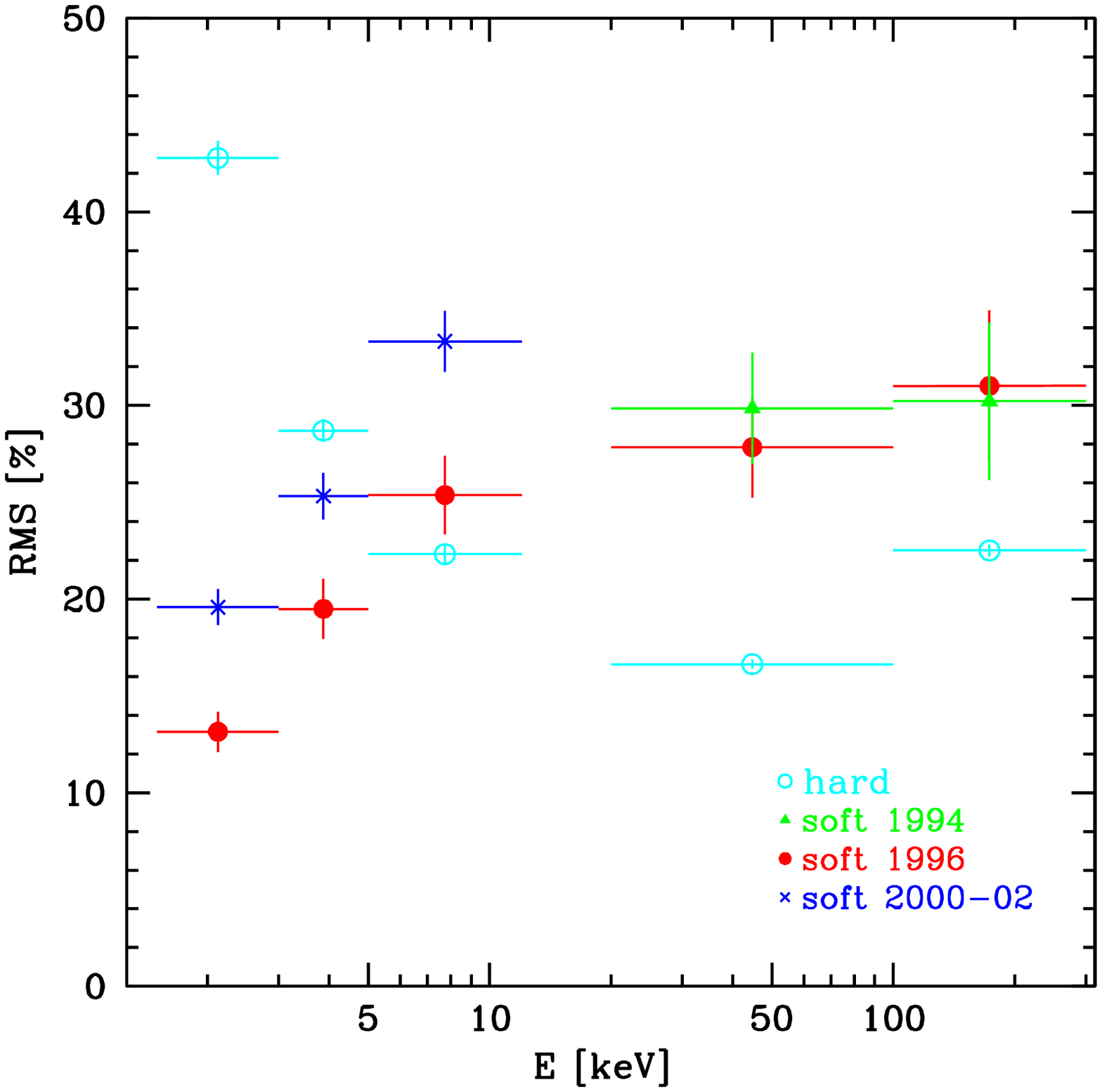}
\caption{The relative rms variability as a function of energy in the hard state
and in the 1994, 1996 and 2000--02 soft states. The respective error bars are
identified by the open circles, filled triangles, filled circles and diagonal
crosses, respectively.
\label{f:rms} }
\end{figure}

An independent way of quantifying the variability of Cygnus X-1 is to consider
the fractional variability as a function of energy. The intrinsic (i.e., after
relatively small corrections for the measurement errors) root mean-square (rms)
variability based on one-day averages in the entire data sets is 89\%, 48\%,
26\%, 21\%, 30\% in the 1.5--3, 3--5, 5--12, 20--100, and 100--300 keV energy
ranges, respectively (see relatively similar values obtained by Z97 for a period
including the 1996 state transtion). However, this variability includes the
state transitions. As we have found in \S \ref{correlations} above, the spectral
variability indicates different physical mechanisms operating in the two main
states. Thus, we have computed the rms variability separately in the hard state
(defined as above and excluding the phase range of 0.85--0.15 for the ASM data)
and in the 1994, 1996 and 2000--02 soft states. The results are shown in Figure
\ref{f:rms}.

%ASM phase 0.15-0.85 only for all data: 88%, 47%, 26%.

The rms dependence in the hard state is consistent with our finding of a pivot 
point in the 20--100 keV range, where the relative variability has the minimum. 
The maximum variability is in the softest channel, consistent with the 
variability being driven by a variable input of soft seed photons (see \S 
\ref{interpretation}). On the other hand, the dependence in the 1996 soft state 
is consistent with a constant soft component and a variable tail with almost 
constant shape. Also, the 1994 soft state (not observed by the ASM) has very 
similar values of the rms, confirming our previous finding of the two states 
being very similar to each other. On the other hand, the 2000--02 soft states 
show an offset of the rms values with respect to those of the 1996 soft state. 
We have checked that it is not due to varying properties of the relatively large 
number of soft episodes during 2000--02. In particular, the rms values for only 
the most recent soft state (MJD $>52250$) are even larger than those shown on 
Figure \ref{f:rms}. A possible explanation for the offset is a decrease of the 
characteristic energy of the constant soft component (see \S \ref{soft02}), 
which then moves the rms pattern to lower energies. In addition, the rms of the 
high-energy tail in the recent soft states appears higher than that in the 1996 
soft state.

In the hard state, the shown values of rms correspond to the frequency range of
$\sim 10^{-8}$--$10^{-5}$ Hz. It is highly interesting to compare the power in
the variability on those long time scales with that corresponding to short time
scales. The rms values corresponding to seven \xte/PCA observations in the hard
state are given by Lin et al.\ (2000). They obtain the rms for 0.002--10 Hz in
the 2--50 keV energy range of $\sim 20$--40\% (with a relatively flat energy
dependence, consistent with the variability at a constant spectral shape), i.e.,
values very similar to those obtained by us. On the other other hand, the
high-frequency hard-state power spectrum per log of frequency of Cygnus X-1 has
an overall maximum (usually with two nearby peaks) in the $\sim 0.1$--10 Hz
range and it declines fast both towards lower and higher frequencies (e.g.,
Gilfanov, Churazov, \& Revnivtsev 1999). This implies that the broad-band
hard-state spectrum of Cygnus X-1 has at least two maxima with comparable power,
one at $\sim 1$ Hz, and one somewhere below $10^{-5}$ Hz. 

\section{Pointed Observations}
\label{pointed}

Further insight into the character of spectral variability of Cygnus X-1 can be
gained from comparing spectra from pointed observations taken at different
times. Here we discuss information that can be derived from spectral variability
between those observations and luminosities in different states.

\subsection{Spectra}
\label{spectra}

\begin{figure*}[t!]
\epsscale{1.6}\plotone{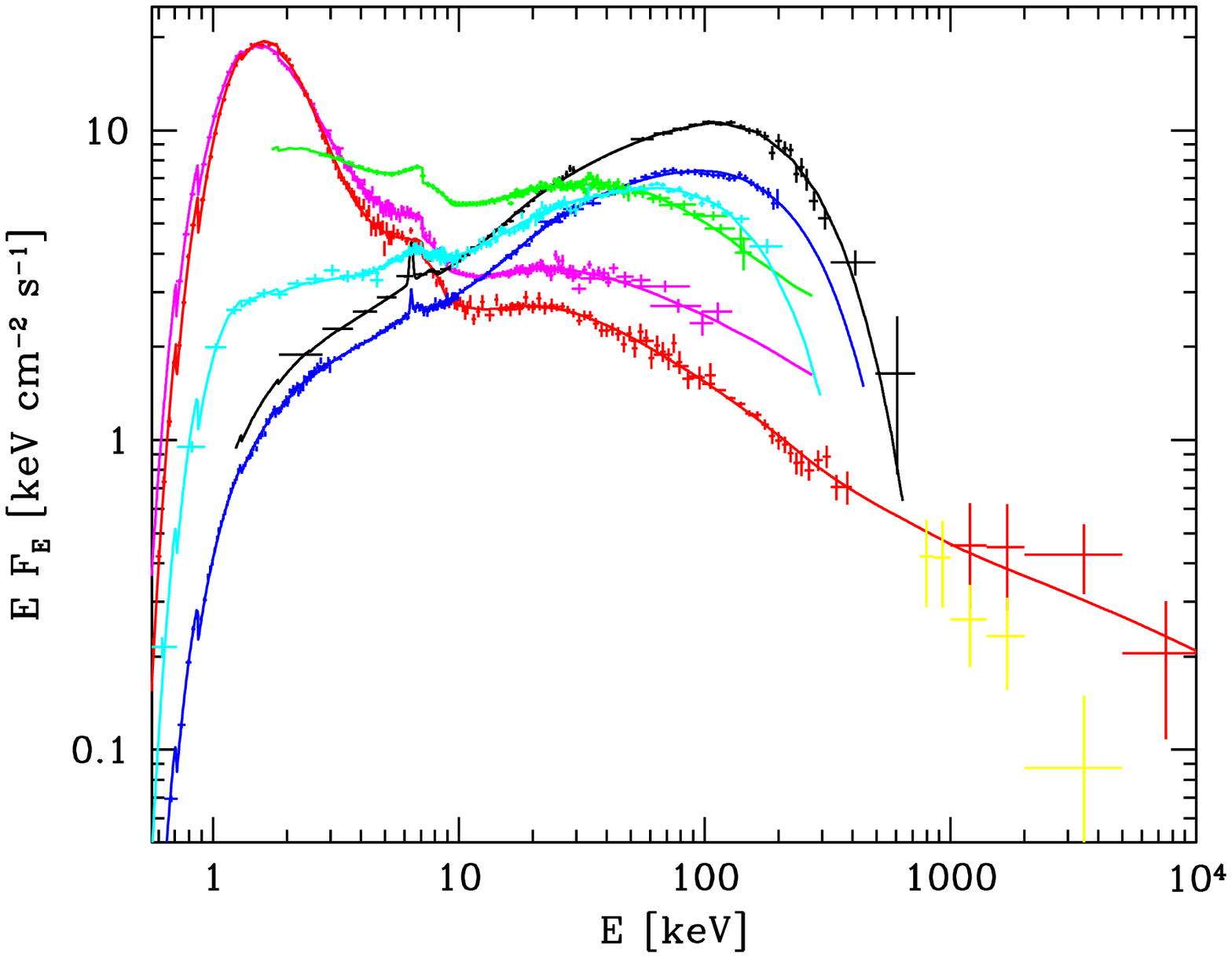}
\caption{Example broad-band spectra from pointed observations. The hard state:
symbols in black: the \ginga-OSSE spectrum from 1991 June 6, no.\ 2 in G97;
blue: the \sax\/ spectrum from 1998 May 3--4 (D01); cyan: the \sax\/ spectrum
from 1996 September 12 (F01). The soft and intermediate states: red: the \sax\/
spectrum from 1996 June 22 (F01) shown together with a \gro/OSSE and COMPTEL
spectra from 1996 June 14--25 (M02); magenta: the \asca-\xte\/ spectrum of 1996
May 30--31 (G99); green: the \xte\/ spectrum of 1996 May 23  (G99). Table 1
gives the detectors to which the spectra are normalized.  The solid curves
give the best-fit Comptonization models (thermal in the hard state, and hybrid,
thermal-nonthermal, in the other states, see \S \ref{interpretation}). For the
\xte\/ spectra, we have updated the PCA response to that of 2002 February. The
symbols in yellow show the average hard-state 0.75--5 MeV spectrum observed by
\gro/COMPTEL (M02).
\label{f:spectra} }
\end{figure*}

Figure \ref{f:spectra} shows three broad-band spectra in the hard state,
two in the soft state, and one in the intermediate state. The triangles and
circles in Figures \ref{a_flux_index}--\ref{a_b_i_i} correspond to four of
those pointed spectra (1996 May 30--31, June 22, September 12, 1998 May 3--4).

In the hard state, we see two patterns of spectral variability. First, the 1991
June (black)  and 1998 May (blue) spectra are almost parallel to each other, and
vary only in the normalization being different by a factor of $\sim 1.5$. (Note
that this 1991 spectrum is thus at about highest daily flux observed by BATSE.)
Such a pattern was also seen within the 1991 observation, when the four spectra
taken within one day (with each observation lasting $\sim 2$ hours) varied only
in the normalization within a factor of $\la 2$ (G97). We can identify this type
of variability with the spread in the $F$-$\Gamma$ correlations in \S
\ref{correlations} at constant values of $\Gamma$, e.g., clearly seen in both
the ASM (Fig.\ \ref{a_flux_index}) and BATSE (Fig.\ \ref{b_flux_index}) data.

Second, the 1996 September hard-state spectrum (cyan)  is substantially softer
than the other hard-state spectra and its high-energy cutoff is at a visibly
lower energy. In a thermal Comptonization fit (see \S \ref{interpretation}
below), $kT\simeq 57$ keV whereas $kT\simeq 86$ keV and 77 keV in the 1991 June
6 and 1998 May 3--4 spectra, respectively. The cyan and blue spectra intersect
at $\sim 50$ keV. We can identify this type variability with the hardness-flux
anticorrelation seen in the ASM data (Fig.\ \ref{a_flux_index}), and the
anticorrelation between the soft and hard X-ray fluxes (Fig.\ \ref{a_b_flux}b).

In the soft and intermediate states, we see the tail (in the red, magenta
and green spectra) moves up and down with relatively small changes in the
spectral shape. The soft X-ray peak is almost the same in the two soft-state
spectra (red and magenta), but the form of the intermediate spectrum above 3 keV
implies a decrease of that peak in the intermediate state. Also, when the tail
moves up, the X-ray spectrum hardens. These effects are consistent with the
behavior of the soft and intermediate state ASM spectra (Fig.\
\ref{a_flux_index}).

\subsection{Bolometric fluxes and the nature of the state transitions}
\label{lum}

One of the outstanding puzzles of Cygnus X-1 has been the small difference,
claimed to be $\la 50$--70\% (Z97) between the bolometric fluxes in the hard and
the soft states. This requires fine-tuning of the hard-state accretion rate as
well as fine-tuning of models of the transition (Esin et al.\ 1998). However, we
stress here that the estimate by Z97 was based on their estimates of the effects
of absorption and the bolometric fluxes based solely on the ASM and BATSE data.

To make further progress, we have compiled in Table \ref{t:lum} the intrinsic 
bolometric fluxes of Cygnus X-1 based on a number of highly accurate broad-band 
pointed observations, including those described in \S \ref{spectra}. Those data 
provide much better estimates of the unabsorbed bolometric flux than those 
possible with the ASM-BATSE data. We see that whereas the typical flux in the 
hard state is $\sim 4$--$5\times 10^{-8}$\,erg\,cm$^{-2}$\,s$^{-1}$, it is $\ga 
1.5\times 10^{-7}$\,erg\,cm$^{-2}$\,s$^{-1}$ in the soft state, i.e., $\sim 
3$--4 times {\it higher}. We also note that the two spectra with the highest 
hard-state fluxes correspond to either the peak of intraday variability or a 
relatively soft continuum. Also, the flux in the intermediate state is just 
between those in the two main states. The fact that the total {\it observed\/} 
flux during the 1996 state transition remains almost unchanged (see Fig.\ 
\ref{lc1}), as pointed out by Z97, can be reconciled with the results in Table 
\ref{t:lum} by noting that whereas the effect of absorption on the measured 
bolometric flux is negligible in the hard state (with the spectrum peaking at 
$\sim 100$ keV, Fig.\ \ref{f:spectra}), it becomes crucial in the soft state, 
and very difficult to estimate with the ASM data alone (with the peak of the $E 
F_E$ spectrum at $\sim 1$ keV).

\begin{deluxetable}{ccccccc}
\tablewidth{0pc}
\tabletypesize{\footnotesize}
\tablecolumns{6}
\tablecaption{Bolometric fluxes and corresponding isotropic luminosities}
\tablehead{\colhead{Date}  & \colhead{State} & \colhead{$F$\tablenotemark{a}} &
\colhead{$L$\tablenotemark{b}} & \colhead{Satellite(s)} &
\colhead{Detector\tablenotemark{c} }}
\startdata
1991--1994\tablenotemark{d} & hard & 3.80 & 1.82 & \gro & OSSE \\
1991 June 6   & hard  & 3.26--5.92\tablenotemark{e} & 1.56--2.83 & \ginga-\gro &
OSSE \\
1996 May 23 & intermediate & 9.80 & 4.69 & \xte & PCA \\
1996 May 30--31 & soft & 15.4 & 7.38 & \asca-\xte & PCA \\
1996 June 22   & soft & 15.1 & 7.24 &
\sax-\gro & LECS \\
1996 Sept.\ 12    & hard & 6.27 & 3.00 & \sax & LECS \\
1998 May 3--4   & hard & 4.42 & 2.12 & \sax & MECS \\
1998 Oct.\ 4--5   & hard & 4.40\tablenotemark{f} & 2.11 & \sax & MECS \\
\enddata
\tablenotetext{a}{The unabsorbed bolometric flux of the model in units of
$10^{-8}$\,erg\,cm$^{-2}$\,s$^{-1}$.}
\tablenotetext{b}{The corresponding luminosity assuming isotropy and a distance
of 2 kpc (see G99) in units of $10^{37}$\,erg\,s$^{-1}$.}
\tablenotetext{c}{The detector to which the flux is normalized.}
\tablenotetext{d}{Using the average \gro\/ spectrum of the hard state of M02 and
a broad-band model with a correction ($+7\%$ derived using \sax\/ data) for the
presence of a soft excess.}
\tablenotetext{e}{The flux range during that day in four observations (G97); the
spectrum shown in Fig.\ \ref{f:spectra} corresponds to the highest flux.}
\tablenotetext{f}{C. Done, private communication.}
\label{t:lum}
\end{deluxetable}

This result makes models of the state transitions based on a change of the 
accretion rate, $\dot M$, much more viable than with the previous result of Z97. 
In particular, a relatively large fraction of the accretion flow in the hard 
state advected (e.g., Narayan \& Yi 1995) or forming outflows (e.g., Blandford 
\& Begelman 1999), i.e., not efficiently radiating, becomes now possible.

\section{Theoretical Interpretation}
\label{interpretation}

We interpret here the spectral variability of Cygnus X-1 in terms of 
Comptonization of soft X-ray blackbody seed photons.  We use a Comptonization 
code by Coppi (1992, 1999), with its present {\sc xspec} (Arnaud 1996) version 
({\tt eqpair}) described in G99.  This model was also used to fit both states of 
Cygnus X-1 by Poutanen \& Coppi (1998), and some states of GRS 1915+105 by 
Zdziarski et al.\ (2001).  In general, the electron distribution can be purely 
thermal or hybrid, i.e., Maxwellian at low energies and non-thermal at high 
energies in the presence of an acceleration process.  The electron distribution, 
including the temperature, $kT$, is calculated self-consistently from the 
luminosities in Comptonized photons, $L_{\rm hard}$, and in the blackbody seed 
photons irradiating the cloud, $L_{\rm soft}$.  The plasma optical depth, 
$\tau$, includes a contribution from \ee\ pairs (which may be negligible or 
not). The importance of pair production depends on the ratio of the luminosity 
to the characteristic size, $r$, which is usually expressed in dimensionless 
form as the compactness parameter, $\ell \equiv L\sigma_{\rm T}/(r m_{\rm e} 
c^3)$ (where $\sigma_{\rm T}$ is the Thomson cross section and $m_{\rm e}$ is 
the electron mass).  We also include Compton reflection (Magdziarz \& Zdziarski 
1995) and Fe K fluorescent lines.  We refer the reader to G99 for details.

We fit this model to the spectra from pointed observations (\S \ref{spectra}).
Using the obtained models, we calculate the energy fluxes in the bands
corresponding to those of the ASM and BATSE. The fluxes from the pointed
observations have uncertainties negligible compared to those of the monitoring
instruments and are very weakly dependent on the assumed model, given the
energy resolution of the pointed instruments being much less than the width of
any of the ASM-BATSE energy bands. Then, in order to compare theoretical
variability patterns with those observed, we change some parameters in some of
the best-fit models, and recalculate the fluxes. For the fitted BATSE indices,
we create simulated BATSE data with models of pointed observations and
theoretical models, and fit them with a power law spectrum. We use two BATSE
responses corresponding to the extreme orientations of the detector with respect
to the source and average the resulting indices. However, the differences in the
two values of $\Gamma$ are $\la 0.03$, i.e., negligibly small.

\subsection{The hard state}
\label{hard}

In the hard state, the Comptonizing plasma is nearly thermal (G97; Poutanen
1998; Zdziarski 2000). The electron distribution may contain a weak high-energy
tail, but its contribution to spectral formation below several hundred keV is
negligible (M02). If the compactness is low enough for pair production being
negligible, the main parameters of a model are $\tau$, $L_{\rm hard}$, and
$L_{\rm soft}$.

We have found that we can explain the anticorrelation of the X-ray flux and
hardness (\S \ref{asm}) by a variable luminosity of seed blackbody photons,
$L_{\rm soft}$, irradiating a cloud of thermal plasma with an approximately
constant energy dissipation rate (given by $L_{\rm hard}$) and a constant
$\tau$. Under these conditions, the plasma electron temperature, $kT$,
will adjust itself to the variable seed flux as to satisfy energy balance. Then,
the higher the seed flux, the lower $kT$, and (since a decrease of $kT$ at a
constant $\tau$ corresponds to an increase of $\Gamma$) the softer the X-ray
spectrum. Since the Comptonized spectrum joins at low energies to the peak of
the spectrum of the seed photons, this will also correspond to an increase of
the soft X-ray flux. The pivoting of the spectrum will occur somewhere below 100
keV causing the decrease of the high energy flux above 100 keV.

\begin{figure}[t!]
\epsscale{1.0}
\plotone{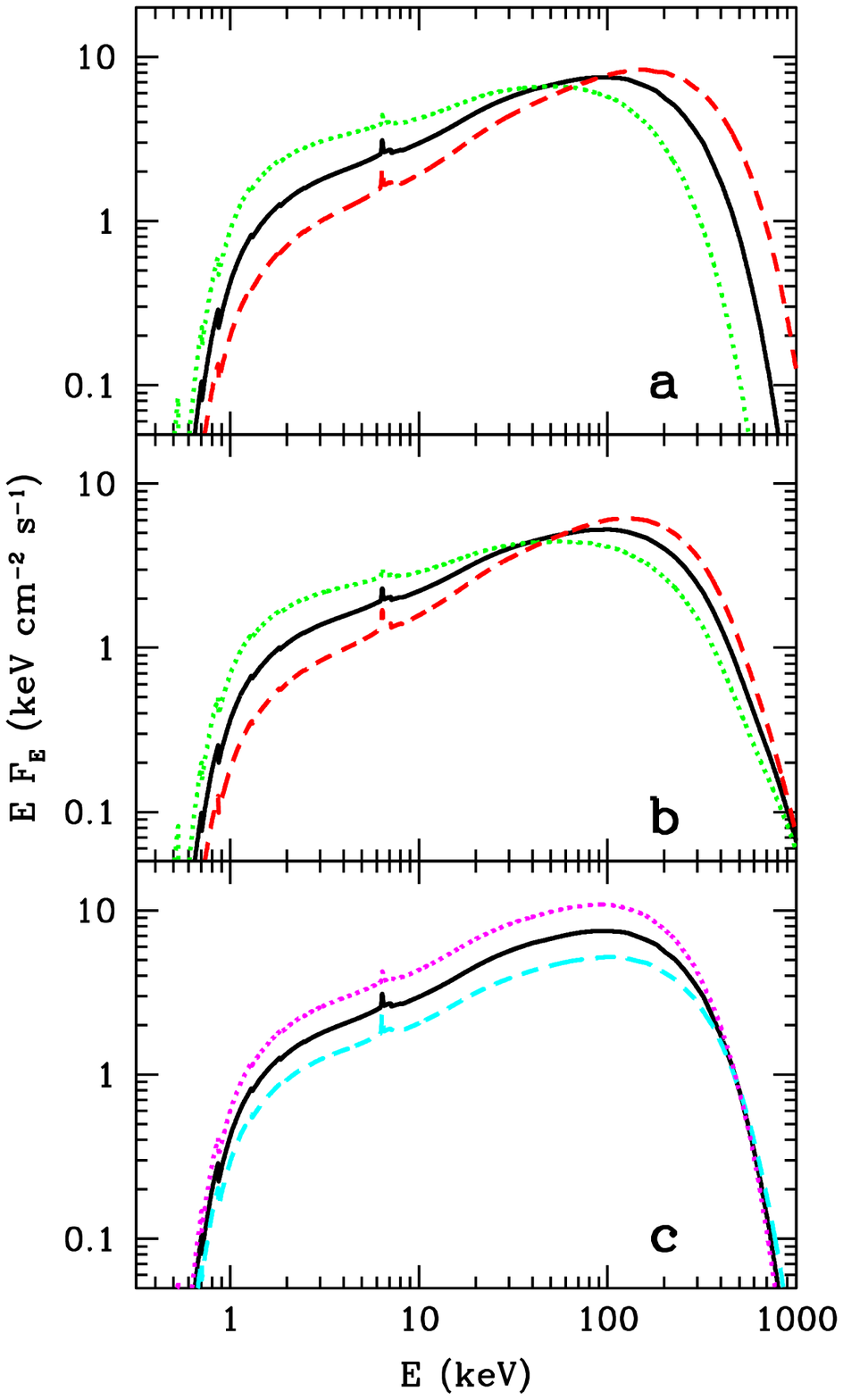}
\caption{The variability patterns (dashed and dotted curves) predicted by our
models in the hard state induced by varying seed soft  photon flux in (a) an
$e^- p$ plasma; (b) \ee\ pair-dominated plasma, and (c) by a change in the
bolometric luminosity due to the change of the local accretion rate. The middle
solid curves correspond to the respective best fits to the 1998 May \sax\/
observation.
\label{patterns_hard} } \end{figure}

We have applied this model to the two hard-state observations of Cygnus X-1 by
\sax\/ (1996 Sept.\ 12, F01; 1998 May 3--4, D01), see \S \ref{spectra}. As
discussed in those papers, Cygnus X-1 shows a soft X-ray excess in addition to
the main thermal Comptonization component. The soft excess is well modeled by an
additional Comptonization region. The seed photons are blackbody at a
temperature of $\sim 0.15$ keV. This model provides a good fit to either of the
data sets, as shown in Figure \ref{f:spectra}. The position of these data in
Figures \ref{a_flux_index}--\ref{a_b_i_i} are shown by the filled triangles and
circles, respectively. We note that there may be some offsets of the \sax\/
points with respect to the corresponding ones from ASM and BATSE due to
calibration uncertainties (see Appendix A) and intraday variability.

\begin{figure*}[t!]
\centerline{\psfig{file=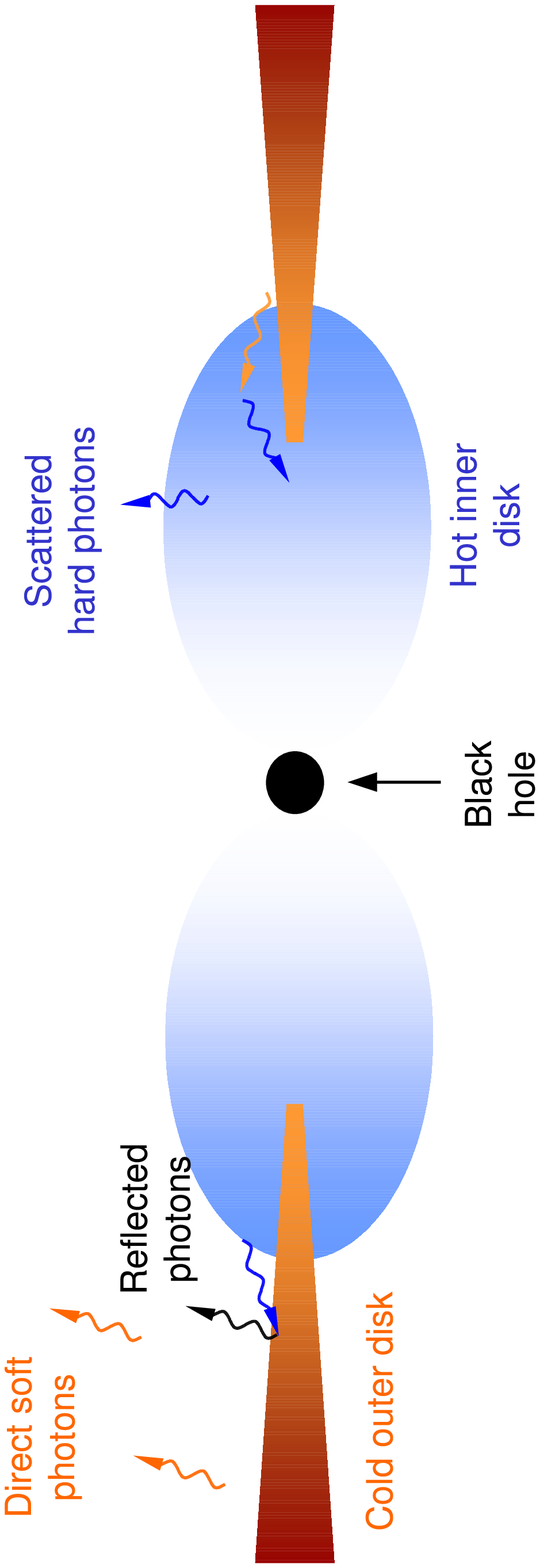,width=14cm,angle=-90}}
\caption{A schematic representation of the likely geometry in the hard state
consisting of a hot inner accretion flow surrounded by optically-thick accretion
disk. The disk is truncated far away from the minimum stable orbit, but it
overlaps with the hot flow. The soft photons emitted by the disk are Compton
upscattered in the hot flow, and emission from the hot flow is partly
Compton-reflected from the disk. The main long-term variability of Cygnus X-1
appears to be due to the variable truncation radius, causing changes in the
hardness of the X-ray spectrum. In a second pattern, the local accretion rate
through the flow changes while the geometry remains constant.
\label{geo_hard} } \end{figure*}

We then vary only $L_{\rm soft}$ for each of the Comptonization regions,
increasing and decreasing it by a factor of 2, and keep the compactness at a low
enough value for pair production to be negligible. We do not adjust the model
normalization. The predicted spectra for the 1998 May observation are shown in
Figure \ref{patterns_hard}a. We see the pivot point at $\sim 50$--90 keV. The
resulting predictions for the fluxes and the spectral hardness are shown in
Figures \ref{a_flux_index}--\ref{a_b_i_i} by the solid black lines originating
at the filled symbols (corresponding to the actual observations). In the case of
the 1996 September observation, we show only the effect of the decreasing
$L_{\rm soft}$ (i.e., hardening of the X-ray spectrum) since the observed
spectrum is already rather soft, see Figure \ref{f:spectra}. Given the
simplicity of our assumption, the  agreement with the observed spectral
variability is rather good, yielding  support to the picture of the X-ray
variability driven by a variable flux of  irradiating soft X-ray photons.
A possible accretion geometry corresponding to this variability pattern
is shown in Figure \ref{geo_hard} (see, e.g. Poutanen, Krolik, \& Ryde 1997;
Esin et al.\ 1998, Zdziarski 1998), and its discussion in given in \S
\ref{discussion}.

The above model is the same as the one used to  explain the X-ray
spectral variability of 3C 120 by Zdziarski \& Grandi (2001). Note that the
pivot obtained for 3C 120 is at $\sim 7$--10 keV, which difference is due to
both the power-law component of 3C 120, $\Gamma\simeq 1.9$, softer than
$\Gamma\sim 1.7$ typical for the hard state of Cygnus X-1, and the lower energy
of the seed photons in AGNs than in black-hole binaries. Specifically, UV seed
blackbody photons with a temperature of 10 eV were assumed for 3C 120, compared
to $\sim 150$ eV fitted for the hard state of Cygnus X-1.

We note that an increase/decrease of $\Gamma$ in this model corresponds to a 
substantial decrease/increase of $kT$, in the $\sim 60$--100 keV range in the 
case shown in Figure \ref{patterns_hard}a. Such an effect is seen when we 
compare the two hard-state \sax\/ spectra. The softer and the harder one have 
$kT\simeq 57$ keV and 77 keV, respectively. However, we do not know how 
universal this correlation is, and the BATSE data are of insufficient accuracy 
to test this effect.

On the other hand, the hot Comptonizing plasma may be \ee\ pair dominated. If it 
is mostly thermal, no signature of annihilation will be visible in the spectra 
(Zdziarski 1986; Macio{\l}ek-Nied\'zwiecki, Zdziarski, \& Coppi 1995). Thus, we 
cannot detect the presence of pairs just by the spectral shape. However, their 
presence will have an effect on the spectral variability. Namely, a variable 
$L_{\rm soft}$ will cause changes in the pair production rate causing changes in 
the total Thomson optical depth, $\tau$. Pairs act as a thermostat (e.g., 
Malzac, Beloborodov, \& Poutanen 2001). We have fitted a pair-dominated model 
(achieved at a high enough compactness, $\ell\simeq 500$ and assuming a 
nonthermal fraction of 5\% consistent with the results of M02) to the 1998 May 
data, and then changed $L_{\rm soft}$ by a factor of 2 as in the case of $e^- p$ 
plasma. Indeed, the equilibrium $kT$ stayed almost constant at 72 keV in all 
three cases.  The resulting variability pattern is shown in Figure 
\ref{patterns_hard}b. We see that the obtained spectra indeed show no 
annihilation features (while pair annihilation is included in the {\tt eqpair} 
model). The predictions of this model are shown by the dotted lines connected to 
the filled circle in Figures \ref{a_flux_index}--\ref{a_b_i_i}. We see that the 
present data are not sensitive enough to determine which of the two cases is 
preferred.

As we discussed in \S \ref{correlations}, there has to be another variability
pattern in the hard state when the total luminosity varies with a more or less
constant spectral shape. If the X-rays in Cygnus X-1 are produced in a hot inner
accretion flow (as in Fig.\ \ref{geo_hard}), a higher accretion rate (and higher
luminosity) corresponds to a higher optical depth, $\tau$,  of the hot disk. If
the geometry of the flow does not change, i.e., the feedback from the cold disk
is constant, the seed photon luminosity (from the outer cool disk) is linearly
proportional to the total luminosity. The resulting spectrum is then almost
independent of luminosity, except that the high energy cutoff decreases with an
increase of $\tau$ since the electron temperature will be adjusted to keep the
Comptonization parameter $y\equiv 4 \tau kT/m_{\rm e} c^2$ constant (see e.g.\
Poutanen 1998 for a review).

We simulate this variability pattern by increasing and decreasing the total
luminosity by $\sqrt{2}$ starting from the same \sax\ spectra of Cygnus X-1 and,
at the same time, adjusting $\tau$ (with no pairs) according to predictions of
models of hot accretion disks.  If $y=$ constant, $\tau\propto L^{2/7}$ and
$L^{1/6}$ in the advection and cooling dominated cases (Zdziarski 1998).  Here
we assumed the former.  A decrease of $kT$ with increasing $L$ at a constant
$\Gamma$ (since $y$ is constant) is consistent with the four data sets of G97,
as well as it has been observed in two bright states of the black hole binary GX
339--4 (Zdziarski 2000; Wardzi\'nski et al.\ 2002).  As we noted above this
variability pattern should have a smaller amplitude comparing with the first
pattern.  The resulting spectra are shown in Figure \ref{patterns_hard}c. The
observed spectra that show similar variability are the hard state spectra (the
black and blue ones) in Figure \ref{f:spectra}. The resulting predictions for
the fluxes and the spectral slopes are shown in Figures
\ref{a_flux_index}--\ref{a_b_i_i} by the dashed lines connected to the filled
circle.  We can see that the two patterns describe well the full range of
observed correlations.

We note that these two variability patterns give opposite predictions for the
dependence of $kT$ on the flux at $\sim 200$ keV. In the case of varying $L_{\rm
soft}$, $kT$ either increases or remains constant when flux increases, and
the opposite is true for the case of varying the total $L$.

\subsection{The 1996 soft state}
\label{soft96}

\begin{figure}[t!]
\epsscale{1.0}
\plotone{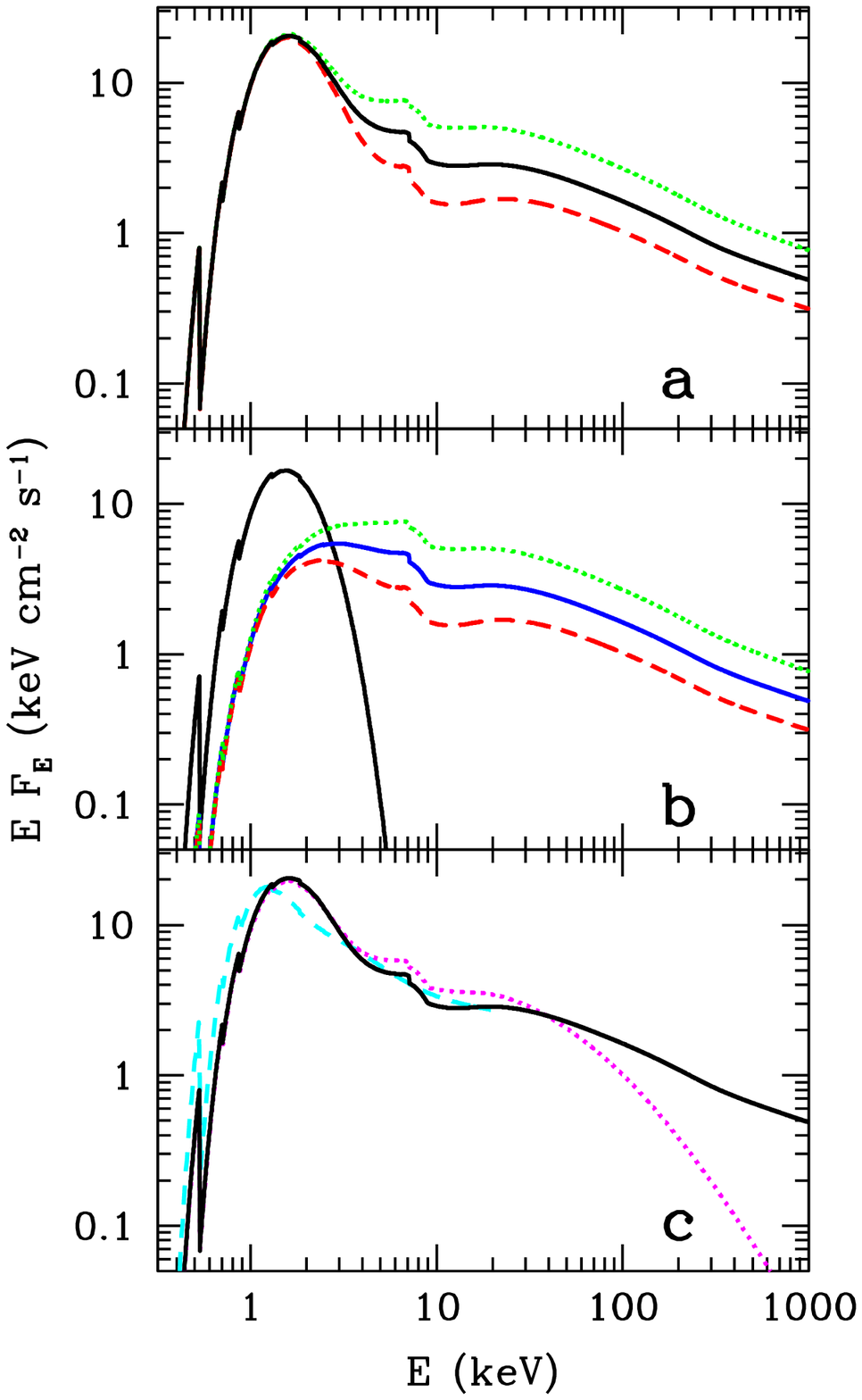}
\caption{(a) The primary variability pattern for  the soft state. The dotted
and dashed curves show the variability induced by varying hard luminosity while
keeping the seed luminosity constant. (b) The constant blackbody and variable
tail components shown separately. (c) The dotted curve illustrates the
secondary variability pattern, due to the maximum Lorentz factor of the
accelerated electrons being reduced from $\gamma_{\rm max}=10^3$ (assumed for
the solid curve) to 10. The dashed curve shows the effect of decreasing the disk
temperature, which can explain the difference in the 2000--02 soft states with
respect to that of 1996. The solid curve in all panels correspond to the best
fit to the 1996 June \sax-\gro\/ spectrum.
\label{patterns_soft} } \end{figure}

For the 1996 soft state, we use the \asca-\xte\/ observation (G99) and the
\sax-\gro\/ one (F01; M02). In those papers, the data were already fitted with
the {\tt eqpair} model, in which a hybrid, thermal-nonthermal plasma Comptonizes
seed, blackbody photons coming from an accretion disk (with the maximum
blackbody temperature of $kT_{\rm soft}$. Now a substantial part ($\sim 0.7$ in
those two observations) of the dissipation in the Comptonizing cloud goes into
particle acceleration, in addition to thermal heating (see G99 for details). The
points corresponding to those observations are shown by the open triangles and
circles, respectively, in Figures \ref{a_flux_index}--\ref{a_b_i_i}.

\begin{figure*}[t!]
\centerline{\psfig{file=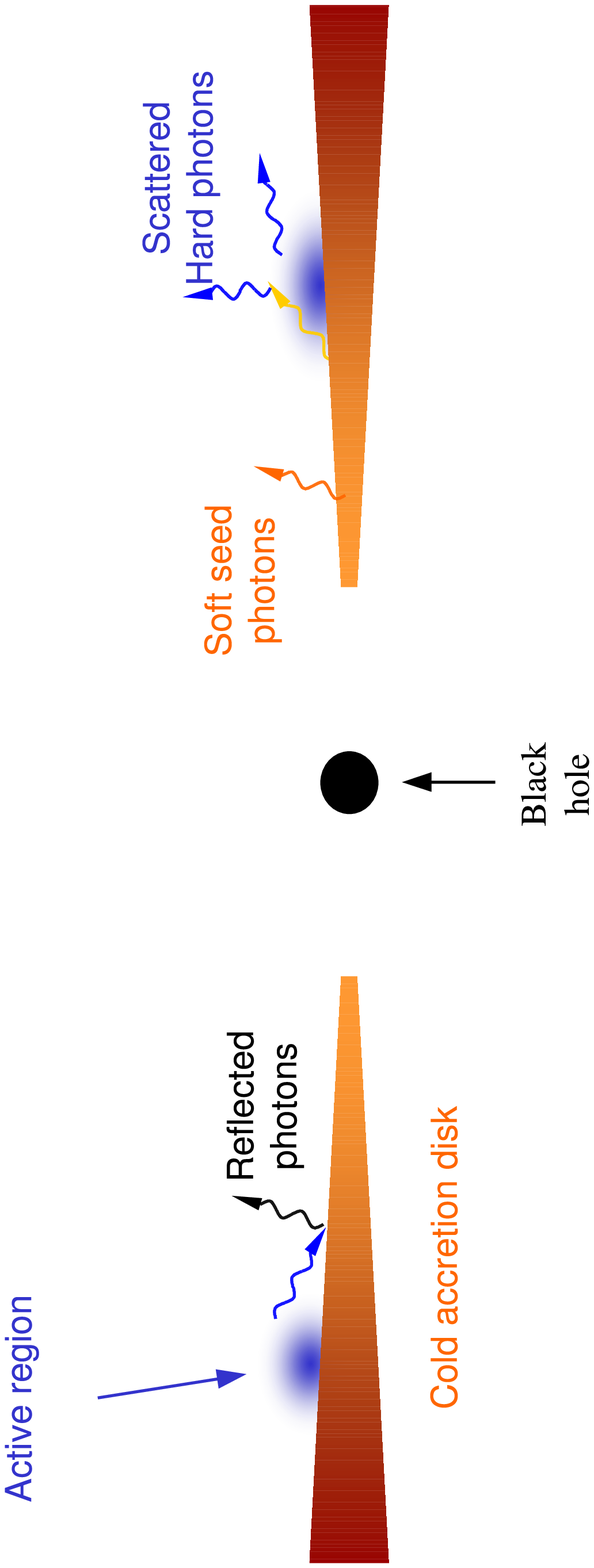,width=14cm,angle=-90}}
\caption{A schematic representation of the likely geometry in the soft state
consisting of flares/active regions above an optically-thick accretion disk
extending close to the minimum stable orbit. The soft photons emitted by the
disk are Compton upscattered in the flares, and emission from the flares is
partly Compton-reflected from the disk. In the case of Cygnus X-1, the disk is
stable whereas the power supplied to the flares is variable during the duration
of the soft state.
\label{geo_soft} } \end{figure*}

Then, we vary $L_{\rm hard}$ by a factor of 1.5 up and down while keeping
$L_{\rm soft}$ constant. The predicted spectra are shown in Figures
\ref{patterns_soft}a, b. This qualitatively explains the spectral variability in
the soft-state region in Figures \ref{a_flux_index}--\ref{a_b_i_i} as shown by
the solid lines extending from the open symbols (corresponding to the actual
observations). This variability pattern  corresponds to a stable optically-thick
accretion disk (Shakura \& Sunyaev 1973) and variable flares on its surface. A
likely corresponding accretion geometry is shown in Figure \ref{geo_soft}, and
more discussion in given in \S \ref{discussion}. The disk in Cygnus X-1 is
thermally and  viscously stable (G99), supporting this picture. The model of
constant disk and a variable tail component has also been found by Churazov,
Gilfanov, \& Revnivtsev (2001) to fit \pca\/ data for Cygnus X-1.

However, the above variability pattern of variable $L_{\rm hard}$ cannot explain 
the correlation of the BATSE spectral slope with the 100--300 keV flux (Figs.\ 
\ref{lc_batse} and \ref{b_flux_index}b) as well as substantial variability of 
that slope (Fig.\ \ref{a_b_index}a).  We note that a softening of the $\sim 
50$--500 keV spectrum with decreasing flux in the same energy range has also 
been observed by OSSE between two observations in the soft states in 1994 and 
1996 (G99).

We have tested a number of models to explain this secondary variability pattern,
with variable either the power-law index of the accelerated electrons, the
fraction of the energy which goes to the acceleration, and the variable maximum
Lorentz factor, $\gamma_{\rm max}$, of the accelerated electrons. We have found
the first pattern does not provide a good fit to the data at all, whereas the
last one provides the best fit. Changes of $\gamma_{\rm max}$ have little effect
on the slope or the flux at low energies (in the ASM range), but markedly affect
the BATSE fluxes and the slope. Starting from the best-fit model for the 1996
June data, we decrease $\gamma_{\rm max}$ from $10^3$ to 10. The resulting
spectrum is shown by the dotted curve in Figure \ref{patterns_soft}c.  The
corresponding predictions for the correlation diagrams are shown for the 1996
June spectrum only by the dotted lines in Figures
\ref{a_flux_index}--\ref{a_b_i_i}.

\subsection{The 2000--02 soft states}
\label{soft02}

As shown in \S \ref{asm} and \ref{rms}, the 2000--02 soft states show different
spectral and timing properties from those of the 1996 soft state. In particular,
Figure \ref{f:rms} shows an offset between the two ASM rms dependencies. In the
framework of the above soft-state model (\S \ref{soft96}) consisting of two
components, a constant disk soft blackbody and a variable tail, the rms offset
is stronly suggestive of a reduction of the disk color temperature. Then, the
variable tail starts to dominate at correspondingly lower energies. From Figure
\ref{f:rms}, we estimate the reduction of the characteristic color temperature,
$kT_{\rm soft}$, to be by a factor of $\sim 1.5$--2.

This inference is, in fact, supported by the ASM colors. As shown in Figures
\ref{a_flux_index} and \ref{a_c_c}, the later soft states have harder 1.5--5 keV
spectra, as well as of a lower flux at a given hardness. This is consistent with
a decrease of $kT_{\rm soft}$, in which case the 1.5--5 keV range has now a
larger contribution from the tail, which has a harder spectrum than the
high-energy cutoff of the disk blackbody, dominant in 1996 up to $\sim 3$ keV
(see Fig.\ \ref{patterns_soft}b).

We have modeled this behaviour using the {\tt eqpair} model. Starting from the
best-fit model for the 1996 June data, we decreased $kT_{\rm soft}$ from 370 eV
(the best fit value) to 200 eV. This, indeed, has led to a hardening of the
1.5--5 keV spectrum. However, we also found that the 3--12 keV hardens as well,
contrary to the data. This implies some change of the parameters of the
Comptonizing plasma in addition to a reduction of $kT_{\rm soft}$.
Unfortunately, the ASM data alone hardly constrain the shape of the
Comptonization spectrum. Here, we found we can account for the ASM data by a
purely phenomenological assumption of setting the reflection fraction to zero.
We stress, however, that the actual character of the change of the hot plasma
can only be determined by broad-band (from soft to hard X-rays) pointed
observations (which were not available to us). E.g, the power-law index of the
accelerated electrons, the fraction of the energy which goes to the
acceleration, or $\gamma_{\rm max}$ could have changed.

The resulting effect on the 1.5--5 and 3--12 keV slopes is shown by the dashed
line originating at the open circle in Figures \ref{a_flux_index}--\ref{a_c_c}.
We see that indeed we can qualitatively account for the changes of the color and
fluxes from the 1996 soft state to the 2000--02 ones. The corresponding spectrum
is shown by the dashed curve in Figure \ref{patterns_soft}c. We show this
spectrum only for energies $\la 20$ keV, since we cannot make any predictions
for higher energies based on the ASM data alone.

Note that we adjusted the normalization of the spectrum to match the data. We
also note the decrease of the disk temperature is not accompanied by a
correspondingly large decrease  in the luminosity, as seen, e.g., in Figures
\ref{lc1} and \ref{lc_asm}. This is possible if the temperature decrease was not
caused by an increasing accretion rate but by a change of some physical
conditions of the disk causing the color correction (the ratio of the color
temperature to the effective one) to decrease. Any more detailed explanation of
the changes requires pointed observations in the soft X-ray range.

The spectral variability {\it within\/} the 2000--02 soft states appears to
follow the primary pattern found for the 1996 soft state, i.e., changes of
$L_{\rm hard}$ at a constant $L_{\rm soft}$. We do not show this pattern
separately in Figures \ref{a_flux_index}--\ref{a_c_c} for the sake of clarity.

\subsection{Discussion}
\label{discussion}

\begin{deluxetable}{cccc}
\tabletypesize{\footnotesize}
\tablewidth{0pc}
\tablecaption{Summary of spectral variability of Cygnus X-1}
\tablehead{\colhead{State} & \colhead{Variability pattern} & \colhead{Model} &
\colhead{Likely physical cause}}
\startdata
Hard & Variable $\Gamma$ with &
Variable $L_{\rm soft}$,  & Hot inner flow
irradiated \\
& a pivot at $\sim 50$ keV  & constant $L_{\rm hard}$ & by a variable outer disk
\\
\\
Hard & Variable amplitude at  &
Variable total $L$ & Variable local $\dot M$ \\
& $\sim$ constant shape & &  \\
\\
Soft & Variable tail on top of &
Constant $L_{\rm soft}$,  & Nonthermal flares
\\
& a constant soft component & variable $L_{\rm hard}$ & above the inner disk \\
\\
Soft & Softening at $\ga 50$ keV &
Less acceleration & Unknown physics \\
& at low fluxes & at low $L_{\rm hard}$ &  of the flares \\
\enddata
\label{t:var}
\end{deluxetable}

Table 2 summarizes our findings above. We stress that the primary variability
patterns in the hard and soft states are modelled by {\it opposite\/} behaviors
of the hard and soft luminosities. This argues agains spectral models of the
state transition related to a quantitative change of one parameter while keeping
the geometry constant, e.g., variations in the relative fraction of the power
dissipated in a disk corona (e.g., Young et al.\ 2001). Although the overall
spectrum in the hard and soft states is indeed hard and soft, respectively,
their spectral variability patterns are completely different.

In the soft state, the spectral variability is indeed fully consistent with the 
presence of flares (forming a corona) on the surface of an accretion disk, as 
schematically shown in Figure \ref{geo_soft}. The luminosity in the flares is 
strongly variable while that in the underlying disk is approximately constant. 
On the other hand, it is likely that the power in the flares is derived from 
dissipation in the underlying disk. In that case, $L_{\rm hard}+L_{\rm soft}$ 
would be approximately constant rather than just $L_{\rm soft}$. We have 
modelled this variability pattern as well, and have found it completely 
consistent with the data but barely distinguishable from the one for constant 
$L_{\rm soft}$ shown in Figures \ref{a_flux_index}--\ref{a_b_i_i}. Thus, we do 
not show it for the sake of simplicity. The 1996 and 2000--02 soft states both 
follow the variability pattern driven by changes of $L_{\rm hard}$, but the 
latter show the disk blackbody temperature lower than that of the former (see 
\S \ref{soft02}).

In contrast, the primary variability in the hard state corresponds to a variable
flux in the seed photons while keeping the power supplied to electrons in the
hot plasma approximately constant. In our opinion, the most likely geometry
appears to be a hot inner flow surrounded by an overlapping thin,
optically-thick, accretion disk supplying the soft photons, as schematically
shown in Figure \ref{geo_hard}. Changes in the inner radius of the disk cause
then variable flux of seed photons. The seed photons are likely to be mostly due
to reprocessing of the hard radiation. As noted in \S \ref{correlations}, this
variability takes place on rather long time scales, which are apparently needed
to change the outer disk. On the other hand, changes in the local $\dot M$ in
the hot disk cause the overall spectrum to move up and down, but with a constant
shape determined by the approximately constant geometry.

An increase in the global $\dot M$ then causes a decrease in the inner radius of
the outer disk, which is compatible with some physical scenarios (e.g., Esin et
al.\ 1998). The data do allow a modest increase of $L_{\rm hard}$ (instead of
its constancy) with increasing $\dot M$, but the present data are not sensitive
enough to test it. When the thin disk moves close to the minimum stable orbit,
the hot accretion flow collapses and is replaced by the active regions above the
thin disk.

On the other hand, it is also possible that the hot plasma forms active coronal
regions above a disk also in the hard state. The variable $L_{\rm soft}$ can be
then related to changes in the bulk velocity of the active regions/flares
(Beloborodov 1999; Malzac et al. 2001) and a constant power dissipated in the
flares causes then the constancy of $L_{\rm hard}$. However, the reversal of
this variability pattern in the soft state needs then to be explained.

We note that, for the sake of simplicity, we have neglected the well-known
correlation between the spectral index and Compton reflection (Zdziarski,
Lubi\'nski, \& Smith 1999; Gilfanov et al.\ 1999, 2000). These changes of the
reflected component would have a relatively minor effect on our results. We
stress, however, that the above correlation is completely consistent with the
flux-hardness anticorrelation in the hard state. Namely, the cold medium
providing the soft seed photons is most likely the same one as the reflecting
medium (see Fig.\ \ref{geo_hard}). Then an increase of the flux of soft photons
due to reprocessing of the hard radiation would also be accompanied by an
increase of the reflection strength. Thus, the presence of one of those
correlations makes it likely that the other one is also present.

\section{Conclusions}
\label{conclusions}

We have presented the history of the flux and hardness of Cygnus X-1 from the 
entire course of its monitoring by the BATSE, and up to 2002 June by the ASM. In 
particular, we show in detail the ASM data from the recent soft states (MJD 
51840 till 52431).

We find then numerous correlations between various ASM-BATSE fluxes and indices.
In the hard state, we find the X-ray flux at $<12$ keV and $>100$ keV to be
negatively and positively, respectively, correlated with the hardness in the
same energy range. We find this is compatible with the variable slope of the
overall spectrum and a pivot at $\sim 50$ keV. This behavior is well modeled by
thermal Comptonization in a hot plasma with an approximately constant power
irradiated by a variable flux of soft seed photons.

In all of the soft states, the X-ray flux at $<12$ keV is positively correlated
with the hardness (as earlier found by Wen et al.\ 2001). However, in the case
of the 1996 soft state, the correlation disappears above 20 keV, where the
spectrum undergoes flux changes at an approximately constant shape. This is well
modeled by hybrid, thermal/nonthermal Comptonizationin a hot plasma with a
variable power irradiated by soft seed photons with a constant flux, i.e., the
behavior opposite to that in the hard state. In addition, we find the overall
1.5--300 keV spectrum is moving up and down in the hard state, and the high
energy tail in the soft state softens at low flux levels. The four patterns are
summarized in Table 2.

We also compute the fractional variability amplitude as a function of photon
energy in the two states. In the hard state, it has a minimum at the 20--100 keV
range, consistent with the presence of the spectral pivot at $\sim 50$ keV, and
the maximum at the 1.5--3 keV range, consistent with the variability being
driven by a variable soft flux. In the soft state, the fractional variability is
lowest in the 1.5--3 keV range and increases strongly towards higher energies,
consistent with the presence of a stable soft (disk) component and a variable
high-energy tail.

Our preferred model of the accretion flows in the hard state is an inner hot
flow overlapping with an outer thin accretion disk with a variable inner radius.
That variability is responsible for varying flux of seed photons and changes of
the spectral slope. However, a dynamic corona above a cold disk is also
possible. In both scenarios, variations of the local accretion rate at the
constant geometry are then responsible for the changes of the luminosity
occuring at an approximately constant shape. In the soft state, the data are
fully compatible with variable nonthermal flares occuring above a stable thin
disk.

We also study in detail differences between the 1996 and 2000--02 soft states.
We find the ASM spectra of the latter to be harder in the 1.5--5 keV range and
to have the rms variability significantly higher. Both effects can be
modeled by a decrease of the color disk temperature from the 1996 soft state to
the 2000--02 ones. On the other hand, we find the 1994 soft state (monitored by
BATSE only) to have the same spectral and timing properties at energies $>20$
keV as the 1996 soft state.

Furthermore, we present a compilation of detailed broad-band spectra of Cygnus
X-1 from pointed observations in the hard, soft, and intermediate states. The
spectral variability observed in this data set fully supports our conclusions
based on the ASM-BATSE data. Furthermore, we find the bolometric luminosity in
the spectra from the pointed observations increases by a factor of $\sim 3$--4
from the hard state to the soft state. This amplitude of the changes of $L$
strongly supports models of the state transitions based on a change of the
global $\dot M$.

\acknowledgements This research has been supported by grants from KBN
(5P03D00821, 2P03C00619p1,2) and the Foundation for Polish Science (AAZ). JP and
AAZ acknowledge support from the Royal Swedish Academy of Sciences and the
Polish Academy of Sciences through exchange programs. We thank the ASM and BATSE
teams for their efforts in carrying out the monitoring, C. Done and T. Di Salvo
for providing us with their \sax\/ data, M. Gierli\'nski for making the original
versions of Figures \ref{geo_hard} and \ref{geo_soft}, and G. Wardzi\'nski and
P. Lachowicz for help with some aspects of the data analysis.

\appendix
\section{Comparison with Pointed Observations}

The ASM results are given in counts per second, $R_i$, in the three adjacent
channels, $i=a, \, b,\, c$, which nominal boundaries are 1.5, 3, 5, 12 keV. In
general, conversion of those rates into energy flux units is uncertain, and
depending on the spectral shape and the position of the source on the sky. Here,
we have adopted a simple conversion based on comparison of fluxes from six
pointed observations with the ASM average daily fluxes taken on the same days.
The pointed observations are listed in Table 1, and correspond to MJD 50226
(\xte, G99), 50233--50234 (\asca-\xte, G99), 50256, 50338 (\sax, F01),
50936--50937 (\sax, D01), MJD 51091 (\sax, C. Done, private communication). The
first five of them are also shown in Figure \ref{f:spectra}.

We have then assumed a linear dependence of the energy flux from each channel on
the count rate in the same one and in a neighbouring one. This takes into
account the dependence of the conversion on the hardness of the incident
spectra. We have then least-square fitted the conversion coefficients. In each
of the three fits, we have given the same relative weight to each of the
measurements, in order to avoid the fit being dominated by the data points with
the highest flux (i.e., in the soft state). In the case of the channel $b$, we
have found that allowing the dependence of the conversion on both neighbouring
channel did not improve the fit quality with respect to the dependence on the
count rate in the $a$, $b$ channels only.

\begin{figure}[t!]
\epsscale{0.9}
\plotone{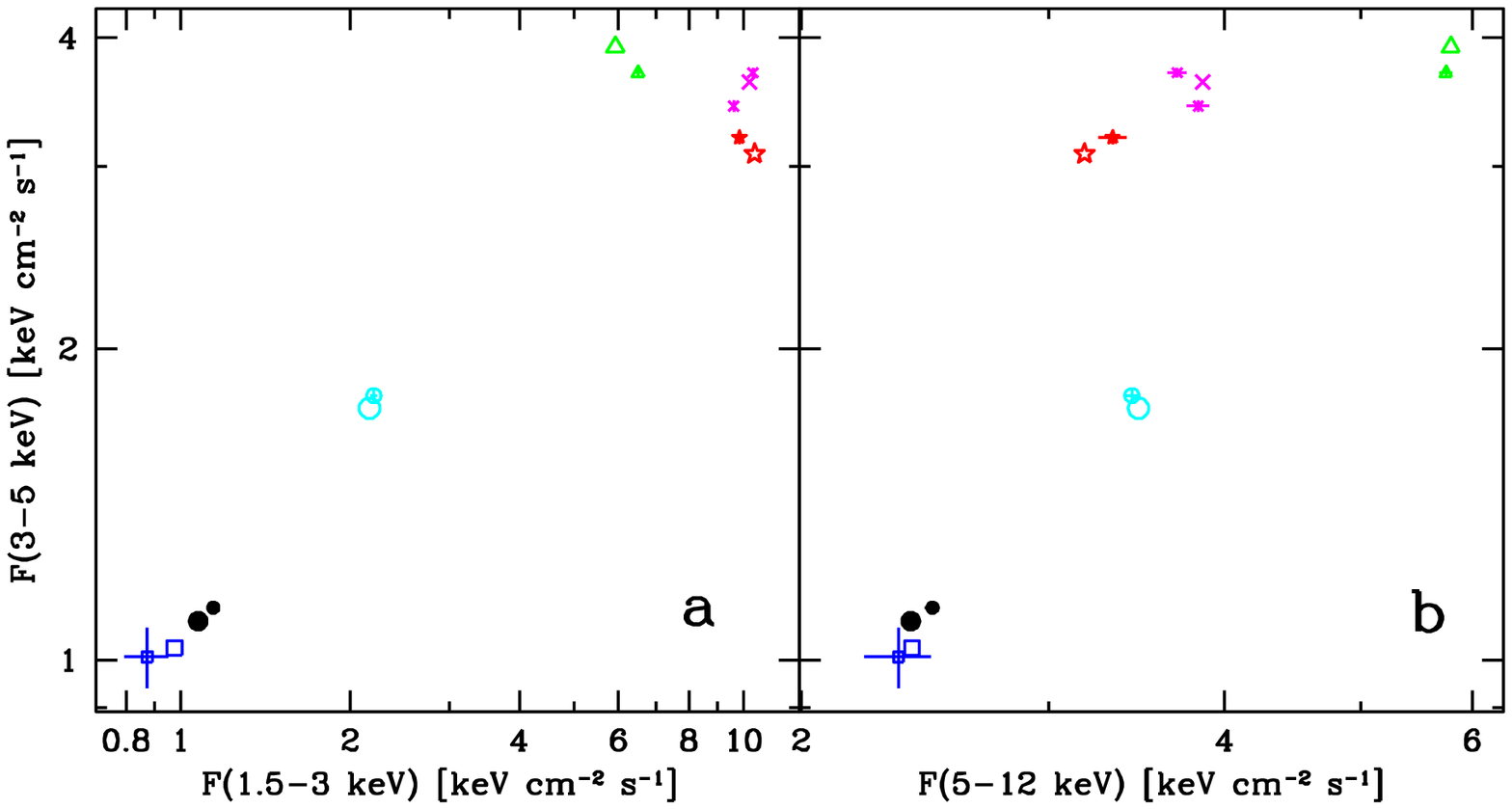}
\caption{Comparison of the fluxes in the three ASM channels (error bars and
small symbols) with those of the corresponding pointed observations (large
symbols). The data corresponding to the six pointed observations correspond to
open triangles, crosses, asterisks, open circles, open squares, and filled
circles, in the temporal order.
\label{asm_compare} }
\end{figure}

In this way, we have obtained,
\begin{equation}
\left( \begin{array}{c}
 F_a  \cr  F_b \cr F_c
\end{array} \right) =
\left( \begin{array}{ccc}
 0.2552  & -0.0324 & 0  \\
-0.0357  &  0.2507 & 0  \\
 0       & -0.0533 & 0.3591
\end{array} \right)
\left( \begin{array}{c}
 R_a  \cr  R_b \cr R_c
\end{array} \right)
\label{eq:conversion}
\end{equation}
where $F_i$ and $R_i$ are in units of keV cm$^{-2}$ s$^{-1}$ and cm$^{-2}$
s$^{-1}$, respectively. We see in Figures \ref{asm_compare}a, b that the
resulting overall agreement is very good, given that some of the differences
have to be systematic, caused by the pointed observations being not taken
exactly on the same times as the ASM observations and by calibration differences
between different instruments.

\begin{figure}[t!]
\epsscale{0.5}
\plotone{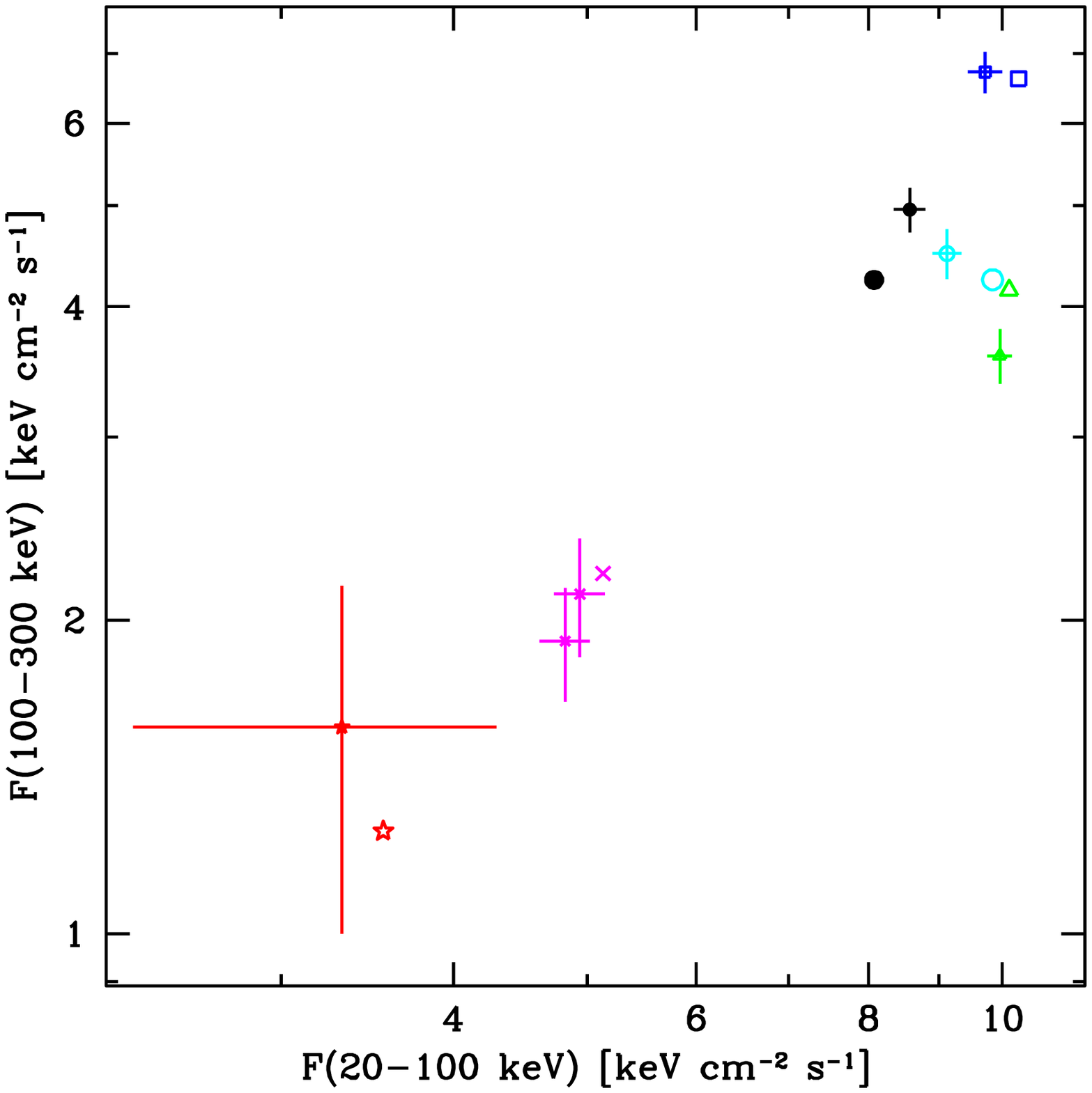}
\caption{Comparison of the fluxes in the two BATSE channels (error bars and
small symbols) with those of the corresponding pointed observations (large
symbols). The data corresponding to the six pointed observations correspond to
open triangles, crosses, asterisks, open circles, open squares, and filled
circles, in the temporal order.
\label{batse_compare} }
\end{figure}

The BATSE fluxes in energy units have been obtained in the way described in \S
\ref{data}. Figure \ref{batse_compare} shows the comparison with the pointed
observations listed above. We see a generally good agreement between the two
data sets.

\end{document}